\documentclass[conference]{IEEEtran}
\usepackage{balance}
\usepackage[utf8]{inputenc}
\usepackage[english]{babel}

\usepackage{etoolbox}
\makeatletter
  \appto\captionsenglish{%
    \renewcommand{\figurename}{FIGURE}%
    \renewcommand{\tablename}{TABLE}%
  }%
  \renewcommand{\figurename}{FIGURE}%
  \renewcommand{\tablename}{TABLE}%
\makeatother

\usepackage[T1]{fontenc} 
\usepackage{microtype} 
\usepackage{csquotes}
\usepackage{amsmath, amssymb, amsfonts}
\usepackage{algorithmic}
\usepackage{graphicx}
\usepackage{xurl}
\usepackage{tabularx}
\usepackage[colorlinks=true, allcolors=blue]{hyperref}
\usepackage{adjustbox} 
\usepackage{float}
\usepackage{multirow}
\usepackage{booktabs}
\usepackage{array}
\usepackage{enumitem}
\graphicspath{{./Figures/}}

\usepackage{cite}  

\newcolumntype{L}[1]{>{\raggedright\arraybackslash}p{#1}} 
\newcolumntype{C}[1]{>{\centering\arraybackslash}p{#1}}   

\begin{document}

\title{
  A Comprehensive Data Description for LoRaWAN Path Loss Measurements in an Indoor Office Setting:\ Effects of Environmental Factors
}

\author{\IEEEauthorblockN{Nahshon Mokua Obiri}
\IEEEauthorblockA{Department of Electrical Engineering and\\
Computer Science\\
University of Siegen, Germany\\
Email: nahshon.obiri@student.uni-siegen.de}
\and
\IEEEauthorblockN{Kristof Van Laerhoven}
\IEEEauthorblockA{Department of Electrical Engineering  and \\
 Computer Science\\
University of Siegen, Germany\\
Email: kvl@eti.uni-siegen.de}}

\maketitle

\begin{abstract}
This paper presents a comprehensive dataset of LoRaWAN technology path loss measurements collected in an indoor office environment, focusing on quantifying the effects of environmental factors on signal propagation. Utilizing a network of six strategically placed LoRaWAN end devices (EDs) and a single indoor gateway (GW) at the University of Siegen’s Hölderlinstraße Campus in the City of Siegen, Germany, we systematically measured signal strength indicators such as the Received Signal Strength Indicator (RSSI) and the Signal-to-Noise Ratio (SNR) under various environmental conditions, including temperature, relative humidity, carbon dioxide (CO\textsubscript{2}) concentration, barometric pressure, and particulate matter levels (PM\textsubscript{2.5}). Our empirical analysis confirms that transient phenomena such as reflections, scattering, interference, occupancy patterns (induced by environmental parameter variations), and furniture rearrangements can alter signal attenuation by as much as \( 10.58\,\mathrm{dB}\), highlighting the dynamic nature of indoor propagation. As an example of how this dataset can be utilized, we tested and evaluated a refined Log-Distance Path Loss and Shadowing Model that integrates both structural obstructions (Multiple Walls) and Environmental Parameters  (LDPLSM-MW-EP). Compared to a baseline model that considers only Multiple Walls (LDPLSM-MW), the enhanced approach reduced the root mean square error (RMSE) from \( 10.58\,\mathrm{dB}\) to \( 8.04\,\mathrm{dB}\) and increased the coefficient of determination (R\textsuperscript{2}) from \( 0.6917\) to \( 0.8222\). By capturing the extra effects of environmental conditions and occupancy dynamics, this improved model provides valuable insights for optimizing power usage and prolonging device battery life, enhancing network reliability in indoor Internet of Things (IoT) deployments, among other applications. This dataset offers a solid foundation for future research and development in indoor wireless communication.  
\end{abstract}


\section{Background}
\label{sec:background}

\subsection{Introduction}
Today, the Internet of Things (IoT) is transforming various industries by enabling a network of interconnected devices that collect and share data. These devices rely on wireless communication technologies to transmit data across different environments \cite{islamFutureIndustrialApplications2024}. Among these technologies, Low-Power Wide-Area Networks (LPWANs) such as Sigfox, Zigbee, Narrowband-IoT (mostly known as NB-IoT), and Long Range Wide Area Network (LoRaWAN) stand out because they offer long-range communication while using very little power \cite{mariniLowPowerWideAreaNetworks2022}.

LoRaWAN, in particular, is becoming the most widely applied low power wide area network (LPWAN) technology due to its open standard, robust security features, and widespread global adoption \cite{garlisiCoexistenceStudyLowPower2023}. It is a communication protocol specifically designed for wireless, battery-operated devices in regional, national, or global networks, addressing key requirements of IoT such as secure bi-directional communication, mobility, and localization services \cite{islamFutureIndustrialApplications2024}. By leveraging Long Range (LoRa) modulation at the physical layer, LoRaWAN enables long-distance communication with minimal power consumption, making it ideal for applications requiring wide coverage and extended battery life \cite{stancoPerformanceIoTLPWAN2022}.

Several features make LoRaWAN suitable for many IoT applications, as surveyed in our recent work \cite{obiriLongRangeWideArea2023}: In smart cities, it enables connectivity for street lighting, parking sensors, and environmental monitoring  \cite{jouhariSurveyScalableLoRaWAN2023}. It supports soil moisture sensing and livestock tracking in agriculture, offering efficient and scalable data transmission for precision farming \cite{sabanSmartAgriculturalSystem2023}. Industrial automation benefits from LoRaWAN's ability to monitor equipment and provide predictive maintenance, enhancing operational efficiency and reducing downtime \cite{garlisiCoexistenceStudyLowPower2023}. In indoor office environments, LoRaWAN can be utilized for building automation systems, enabling energy-efficient control of lighting, heating, ventilation, and air conditioning (HVAC) systems. It also supports occupancy sensing for better space utilization, asset tracking for equipment management, and environmental monitoring to ensure optimal working conditions \cite{langoteSystemIoTAgriculture2023}. Additionally, in our previous work \cite{obiriSurveyLoRaWANIntegratedWearable2024a}, we explored the potential of LoRaWAN in human activity recognition (HAR), demonstrating its efficiency in integrating wearable sensor networks for activities of daily living (ADL), healthcare, localization, and user safety applications.

While LoRaWAN is widely used, understanding how its signals behave inside buildings is essential for optimizing its performance \cite{aksoyComparativeAnalysisEnd2024a, alkhazmiAnalysisRealWorldLoRaWAN2023a}. Indoor environments, such as office spaces, present unique challenges for wireless communication. Walls, floors, furniture, and the movement of people can weaken the signal through attenuation, reflection, and multipath fading, making it harder to maintain a reliable connection. Building materials like concrete, glass, or metal can absorb or reflect radio signals, while wall thickness and metallic structures such as reinforcements or ducts can exacerbate path loss. Electronic devices and dynamic human movement can also create interference and multipath effects, further complicating signal behavior.

Traditional path loss models (PLMs) often have limitations when applied to specific environments like modern office buildings, where diverse materials and layouts can significantly affect signal propagation \cite{wuArtificialNeuralNetwork2020}. Generic models may not accurately predict signal behavior in such complex settings. Given these challenges, empirical measurements become essential to characterize path loss in specific indoor environments accurately \cite{alobaidyWirelessTransmissionsPropagation2022a}. By conducting on-site measurements, we can capture the unique properties of the environment and adjust the models accordingly, allowing for the development of customized PLMs that reflect the actual conditions in which the LoRaWAN network operates.

In the context of LoRaWAN technology, understanding and modeling path loss is crucial due to its reliance on low power and long-range communication \cite{jouhariSurveyScalableLoRaWAN2023}. Accurate PLMs facilitate effective network planning for real-time deployments \cite{saidPerformanceEvaluationLoRaWAN2024a} and enable the optimization of GW placements and transmission parameters to ensure reliable connectivity \cite{alobaidyWirelessTransmissionsPropagation2022a}. Environmental factors such as temperature, relative humidity, particulate matter, and barometric pressure affect signal propagation \cite{ohshimaFieldExperimentsDeveloping2010, deeseLongTermMonitoringSmart2021, zhangImpactPM25Satelliteground2024}. The SNR is a crucial variable influencing link quality and overall network performance \cite{biancoLoRaSystemSearch2021a}. Therefore, incorporating environmental parameters and SNR into PLMs can significantly enhance their accuracy. Recent studies have successfully integrated these variables to improve path loss modeling in outdoor LoRaWAN deployments \cite{gonzalez-palacioLoRaWANPathLoss2023}. However, besides these breakthroughs, accounting for human activities introduces additional complexity for indoor path loss, presenting complex challenges in developing robust PLMs.

\subsection{Research Aim and Contributions}

This study aims to systematically investigate the impact of indoor environmental factors on LoRaWAN signal propagation in a typical office setting. By conducting controlled path loss measurements and analyzing signal strength under varying conditions, we offer data-driven insights for optimizing indoor IoT deployments. It builds directly on our prior survey \cite{obiriSurveyLoRaWANIntegratedWearable2024a}, which outlined key challenges in LoRaWAN-based HAR systems, particularly in dynamic environments. The dataset introduced here is the first of its kind regarding scale, diversity, and environmental granularity. 

To achieve this aim, the paper follows four main interconnected research paths: \textit{\textbf{(i)}} surveying the state-of-the-art in indoor LoRaWAN propagation models; \textit{\textbf{(ii)}} designing and executing a structured, reproducible measurement campaign; \textit{\textbf{(iii)}} analyzing and describing the resulting dataset; and \textit{\textbf{(iv)}} demonstrating practical application through an applied path loss modeling example. Consequently, this paper offers several key contributions that advance the understanding of indoor LoRaWAN network performance, outlined as follows:

\begin{enumerate}[label=(\roman*)]
    \item We comprehensively analyze current state-of-the-art models' literature and approaches for indoor LoRaWAN signal propagation, highlighting their limitations in accounting for environmental factors.
    \item We present a novel dataset of LoRaWAN path loss measurements, comprising \( 1,\!328,\!334 \) fields collected in an indoor office environment. It includes detailed metrics such as the received signal strength indicator (RSSI), SNR, and environmental parameters (temperature, relative humidity, carbon dioxide (CO\textsubscript{2}), barometric pressure, and particulate matter (PM\textsubscript{2.5}) levels), providing a valuable resource for further research and analysis.
    \item We conduct an extensive empirical analysis to quantify the impact of various environmental factors and human activity on LoRaWAN signal strength and quality. Our findings reveal significant changes in signal loss due to these factors, which have been underexplored in previous studies.
    \item We demonstrate how the collected data can be used and validate an improved Log-Distance Path Loss and Shadowing Model (LDPLSM) that incorporates both structural obstructions (walls of different materials) and environmental parameters. Compared to a baseline model that considers only structural factors, the enhanced model yields more accurate predictions of signal attenuation in indoor office environments, reducing the root mean square error (RMSE) from \( 10.58\,\mathrm{dB} \) to \( 8.04\,\mathrm{dB} \) and increasing the coefficient of determination (R\textsuperscript{2}) from \( 0.6917 \) to \( 0.8222 \). 
\end{enumerate}

\subsection{Paper Organization}

The remainder of this paper is organized as follows:
\begin{itemize}
    \item Section \ref{sec: lorawan_technology} provides an overview of LoRaWAN technology and discusses related works in the field. 
    \item Section \ref{sec:related_work} further reviews existing literature and related studies pertinent to our research. 
    \item Section \ref{sec: methodology} details the experimental design and methods of our setup, including the devices used, data collection procedures, and environmental conditions considered. 
    \item In Section \ref{sec: data_description}, we describe the dataset and perform an analysis of the collected data. 
    \item Section \ref{sec: Indoor_Models} covers data preparation and parameter estimation for modeling and evaluation of indoor path loss and shadowing, including both primary and enhanced approaches that account for environmental factors affecting signal propagation.
    \item Section \ref{sec: conclusion} concludes the paper and suggests directions for future research.
\end{itemize}

\subsection{Abbreviations and Acronyms}

Table~\ref{tab:abbreviations} lists the abbreviations and acronyms and their definitions commonly used in this paper.

\begin{table}[hbt!]
\centering
\scriptsize
\caption{List of Abbreviations and Acronyms}
\begin{tabular}{p{60pt}p{155pt}}
\toprule
\textbf{Abbreviation} & \textbf{Definition} \\
\midrule
ADR           & Adaptive Data Rate \\
AFTD          & Adeunis Field Test Device \\
ANN           & Artificial Neural Network \\
AWS           & Amazon Web Services \\
BW            & Bandwidth \\
CO\textsubscript{2}   & Carbon Dioxide \\
CR            & Coding Rate \\
CSS           & Chirp Spread Spectrum \\
ED            & End Device \\
ESP           & Effective Signal Power \\
GW            & Gateway \\
HAR           & Human Activity Recognition \\
I2C           & Inter-Integrated Circuit\\
IoT           & Internet of Things \\
IP            & Internet Protocol\\
ISM           & Industrial, Scientific, and Medical \\
LDPLM         & Log-Distance Path Loss Model \\
LDPLSM         & Log-Distance Path Loss and Shadowing Model \\
LDPLSM-MW      & LDPLSM with Multiple Walls \\
LDPLSM-MW-EP   & LDPLSM with Multiple Walls and Environmental Parameters \\
LoRa          &  Long Range \\
LoRaWAN       & Long Range Wide Area Network \\
LoS           & Line-of-Sight \\
LPWAN         & Low Power Wide Area Network \\
MQTT          & Message Queuing Telemetry Transport \\
NLoS          & Non-Line-of-Sight \\
NP            & Noise Power \\
PER           & Packet Error Rate \\
PDR           & Packet Delivery Ratio \\
PLM           & Path Loss Model \\
PM\textsubscript{2.5} & Particulate Matter less than \( 2.5\,\mu\mathrm{m} \) \\
RMSE          & Root Mean Square Error \\
RSSI          & Received Signal Strength Indicator \\
SF            & Spreading Factor \\
SNR           & Signal-to-Noise Ratio \\
SPI  & Serial Peripheral Interface\\
ToA           & Time on Air \\
TTN           & The Things Network \\
UART  &  Universal Asynchronous Receiver/Transmitter\\
VM            & Virtual Machine \\
\bottomrule
\end{tabular}
\label{tab:abbreviations}
\end{table}

\section{LoRaWAN Technology Overview}
\label{sec: lorawan_technology}

\subsection{Architecture and Network Topology}

\begin{figure}[hbt!]
\centering
\centerline{\includegraphics[width=\columnwidth]{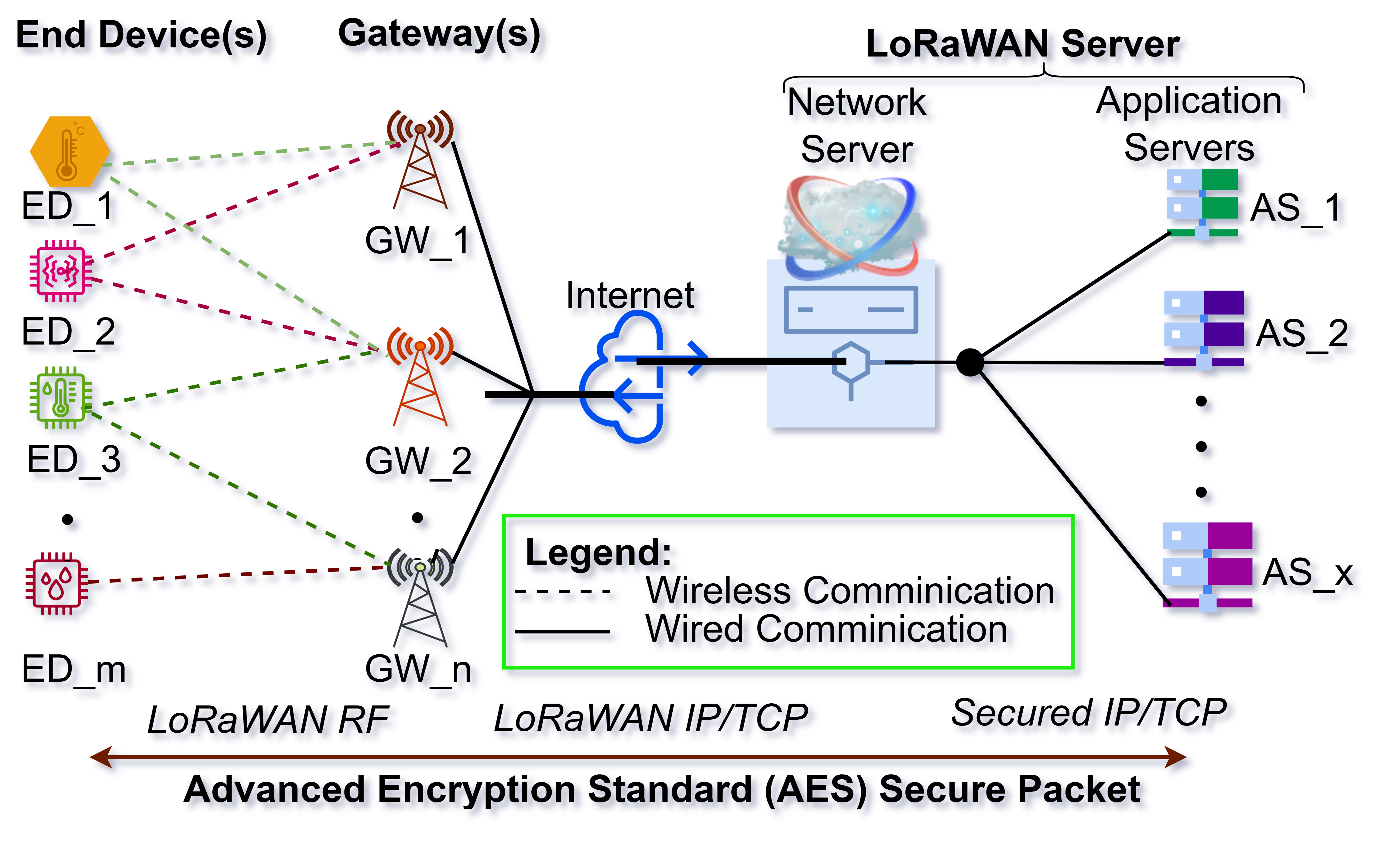}}
\caption{Standard LoRaWAN Architecture: Overview showing end devices (EDs) connecting through gateways (GWs) to the LoRaWAN server (network and application servers) via the Internet.}
\label{fig:LoRaWAN}
\end{figure}

LoRaWAN employs a star-of-stars network topology as shown in Fig.~\ref{fig:LoRaWAN}, where gateways (GWs) relay messages between end devices (EDs) and a central network server \cite{milarokostasComprehensiveStudyLPWANs2023}. EDs are sensors or actuators embedded on microcontrollers with LoRa transceivers communicating with a single or multiple GWs that act as transparent bridges, forwarding messages to the network server via standard internet protocol (IP) connections like Wi-Fi, Ethernet, or cellular networks. To optimize network performance, the network server manages core functions such as data filtering, security checks, and adaptive data rate (ADR) management. The multi-gateway approach enhances network resilience and scalability in this topology, allowing messages to reach the server through different paths \cite{almuhayaSurveyLoRaWANTechnology2022}. Moreover, communication is bidirectional, where uplink transmissions send data from EDs to the network server, while downlink transmissions send commands or acknowledgments from the network server to EDs.

\subsection{Communication Protocols and Data Transmission}

LoRa is a physical layer technology that utilizes chirp spread spectrum (CSS) modulation to enable long-range, low-power wireless communication. At the same time, LoRaWAN operates at the medium access control layer, managing communication protocols and network access \cite{milarokostasComprehensiveStudyLPWANs2023}. This layered approach makes the system ideal for IoT applications \cite{magrinPerformanceAnalysisLoRaWAN2021a}. In CSS modulation, a chirp signal sweeps linearly across a designated bandwidth (BW), increasing or decreasing in frequency over time \cite{milarokostasComprehensiveStudyLPWANs2023}. This technique provides robustness against interference, multipath fading, and Doppler effects, thereby enhancing the reliability of data transmission.

The central carrier frequency \( f_0 \) serves as the midpoint of the chirp's frequency sweep, which spans from \( f_0 - \mathrm{BW}/2 \) to \( f_0 + \mathrm{BW}/2 \). A key parameter in LoRaWAN is the spreading factor (SF), which determines the number of bits per symbol and controls the trade-off between data rate and communication range. SF values range from 6 to 12; increasing the SF lengthens the symbol duration \( T_{\text{symbol}} \), allowing for improved detection of weak signals at the cost of reduced data rates. Equation~(\ref{eq: symbolduration}) provides the calculation for the symbol time \( T_{\text{symbol}} \). It indicates that each increment in $SF$ doubles the symbol time \( T_{\text{symbol}} \), impacting both data rate and energy consumption. Also, longer symbol times decrease the data rate and increase energy usage due to extended transmission durations. In addition to affecting data rate, the SF also influences the ToA, the total duration required to transmit an entire LoRaWAN packet.
\begin{equation}
\label{eq: symbolduration} 
    T_{\mathrm{symbol}}  = \frac{2^{\mathrm{SF}}}{\mathrm{BW}} .
\end{equation}

LoRaWAN incorporates a coding rate (CR) to enhance error correction, which adds redundancy to the transmitted data \cite{jouhariSurveyScalableLoRaWAN2023}. The CR is expressed as $4/(4 + n)$, with $n$ taking values from 1 to 4, corresponding to CRs of 4/5, 4/6, 4/7, and 4/8, respectively. A higher $n$ adds more redundancy bits, improving error correction capabilities but reducing the effective data throughput. SF, BW, and CR influence the data rate \( R_b \) and can be approximated by  (\ref{eq: datarate}) as:
\begin{equation} 
\label{eq: datarate} 
    R_b = \frac{\mathrm{SF} \times \mathrm{BW}}{2^{\mathrm{SF}}} \times \mathrm{CR},
\end{equation}
where $R_b$ represents the bit rate. This relationship shows that the data rate decreases exponentially with increasing SF, while higher BWs and CRs can help mitigate this effect.

Adjusting SF and CR allows devices to optimize communication based on link conditions. Higher SFs and lower CRs improve sensitivity and range but reduce data rates and increase energy consumption. Conversely, lower SFs and higher CRs enable faster data rates and lower energy usage but require stronger signal conditions. By leveraging CSS modulation and carefully selecting parameters like SF, BW, and CR, LoRaWAN achieves robust and scalable communication \cite{milarokostasComprehensiveStudyLPWANs2023}. This modulation strategy enables devices to maintain reliable connections over long distances while conserving energy, which is crucial for battery-powered IoT devices.

\subsection{Effective Signal Power and Noise Level}

In LoRaWAN, the effective signal power (ESP) quantifies the portion of the received power attributable to the actual signal, excluding background noise. This metric provides a clearer understanding of signal strength by isolating it from interference and noise factors. Equation (\ref{eq: esp}) expresses the ESP on a logarithmic scale as given in \cite{tongCitywideLoRaNetwork2024}.
\begin{equation}
\label{eq: esp}
    \mathrm{ESP}_{\mathrm{dBm}} = \mathrm{RSSI}_{\mathrm{dBm}} + \mathrm{SNR}_{\mathrm{dB}} - 10 \times \log_{10} \left( 1 + 10^{0.1 \times \mathrm{SNR}_{\mathrm{dB}}} \right) ,
\end{equation}
where \( \mathrm{ESP}_{\mathrm{dBm}} \) and \( \mathrm{RSSI}_{\mathrm{dBm}} \) (indicating the total power received by the antenna) are in decibel milliwatt (dBm) and \( \mathrm{SNR}_{\mathrm{dB}} \) is given in decibels (dB) units and it reflects the ratio between the desired signal power and the background noise power (NP).

The NP indicates the strength of the background interference affecting the received signal. In the log scale, it is given in  (\ref{eq:noise_power}) \cite{gonzalez-palacioLoRaWANPathLoss2023} as:
\begin{equation}
\label{eq:noise_power}
    P_{n\,\mathrm{dBm}} = \mathrm{RSSI}_{\mathrm{dBm}} - 10 \times \log_{10} \left(1 + 10^{0.1 \times \mathrm{SNR}_{\mathrm{dB}}} \right),
\end{equation}
where \( P_{n\,\mathrm{dBm}} \) denotes the the NP in \(\mathrm{dB}\).

Successful reception of LoRaWAN packets relies on the ESP being above the receiver sensitivity threshold for a given SF, as shown in Table~\ref{tab:SNR_SF_Sensitivity}. Factors like distance, obstacles, and environmental conditions can reduce signal strength below this threshold, leading to communication failures. In indoor environments, multipath fading and walls can cause ESP fluctuations, while electronic interference and signal reflections can increase NP, impacting communication reliability \cite{kufakunesuCollisionAvoidanceAdaptive2024a}. Additionally, interference and collisions occur when multiple devices transmit simultaneously on the same frequency and SF, reducing the effective SNR. If the SNR falls below the required threshold specified in Table~\ref{tab:SNR_SF_Sensitivity}, packets cannot be decoded successfully. Understanding these factors is essential for optimizing node placement and network configuration to ensure robust and reliable data transmission in LoRaWAN networks, thus improving the packet delivery ratio (PDR) in high-interference places of deployment.

\begin{table}[hbt!]
\centering
\caption{Required Signal-to-Noise Ratio (SNR) and Receiver Sensitivity for each Spreading Factor (SF)  \cite{kufakunesuCollisionAvoidanceAdaptive2024a}}
\label{tab:SNR_SF_Sensitivity}
\begin{tabular}{ccc}
\toprule
\textbf{Spreading Factor (SF)} & $\mathbf{SNR}_{\text{req}}$ \textbf{(dB)} & \textbf{Receiver Sensitivity} (dBm) \\
\midrule
SF7  & -7.5  & -123 \\
SF8  & -10.0 & -126 \\
SF9  & -12.5 & -129 \\
SF10 & -15.0 & -132 \\
SF11 & -17.5 & -134.5 \\
SF12 & -20.0 & -137 \\
\bottomrule
\end{tabular}
\end{table}

\subsection{Adaptive Data Rate}

ADR in LoRaWAN is a mechanism designed to optimize ED data transmission settings to enhance network efficiency and extend battery life. By dynamically adjusting both the SF and the transmit power based on link quality, it ensures that each device communicates using the most appropriate settings for its current conditions \cite{milarokostasComprehensiveStudyLPWANs2023}. The ADR process begins when an ED signals its capability to participate in ADR by setting the designated ADR bit to '1' in its uplink frames. The network server then collects link quality metrics from recent uplink messages, explicitly focusing on the SNR.

To assess the link quality, the network server calculates the SNR margin using the maximum SNR from recent uplink messages as given in  (\ref{eq: snr}) \cite{gonzalez-palacioMachineLearningBasedCombinedPath2023a} as:
\begin{equation}
\label{eq: snr}
    \mathrm{SNR}_{\mathrm{margin}} = \max_{i=1}^{N} \mathrm{SNR}_i - \mathrm{SNR}_{\mathrm{req}} - M ,
\end{equation}
where $N$ is the number of recent uplink messages considered (mostly \(20\)), \( \mathrm{SNR}_i \) is the SNR of the \( i \)-th message, \( \mathrm{SNR}_{\mathrm{req}} \) is the required SNR for the current SF as specified in Table~\ref{tab:SNR_SF_Sensitivity}, and $M$ is a fade margin to account for signal variability, typically set to \( 10\,\mathrm{dB} \). Since maximum SNR reflects optimistic conditions and favors higher data rates, it remains common in LoRaWAN implementations. Some studies suggest using average SNR for more robust link estimation \cite{kufakunesuCollisionAvoidanceAdaptive2024a}.

If the calculated \( \mathrm{SNR}_{\mathrm{margin}} \) is greater than zero, the link quality is better than necessary for the current settings. The network server can then increase the data rate by lowering the SF, which reduces airtime and conserves energy. If the \( \mathrm{SNR}_{\mathrm{margin}} \leq 0 \)  and the SF cannot be decreased further (i.e., the minimum SF is already in use), the network server may decide to increase the transmit power of the ED to improve link quality. The new transmit power is calculated as shown in  (\ref{eq: tp}):
\begin{equation} 
\label{eq: tp} 
    P_{\mathrm{new}} = \min\left( P_{\mathrm{current}} + \Delta P, P_{\mathrm{max}} \right) ,
\end{equation}
where $P_{\text{current}}$ is the current transmit power, $\Delta P$ is the power adjustment step, and $P_{\text{max}}$ is the maximum allowed transmit power. Once the optimal SF and transmit power are determined, the network server communicates these settings to the ED, which applies the new settings and acknowledges receipt.

\subsection{Security Mechanisms in LoRaWAN}

Security is a cornerstone of LoRaWAN. The protocol ensures data confidentiality, integrity, and authenticity through encryption and authentication mechanisms based on the Advanced Encryption Standard (AES) \cite{loukilAnalysisLoRaWAN102022}. Separate session keys are used to strengthen security. The Network Session Key (NwkSKey) authenticates devices and ensures message integrity at the network layer. The Application Session Key (AppSKey) encrypts and decrypts application data at the application layer, maintaining confidentiality. This separation of keys protects against unauthorized access and ensures that only authorized devices can participate in the network, reinforcing the overall security framework.

\subsection{LoRaWAN Spectrum Usage}

LoRaWAN operates within the unlicensed industrial, scientific, and medical (ISM) frequency bands, enabling cost-effective deployment without spectrum licensing fees \cite{islamFutureIndustrialApplications2024}. In Europe, it utilizes the \( 863 \) to \( 870\,\mathrm{MHz} \) frequency range, which is segmented into multiple channels typically using a BW of \( 125\,\mathrm{kHz} \) for both uplink and downlink communications. European regulations, such as those specified by the European Telecommunications Standards Institute (ETSI), impose duty cycle limitations to ensure fair spectrum usage; devices are generally restricted to a duty cycle of 1\% or 0.1\%, depending on the sub-band. This constraint means that a device can transmit only for a limited percentage of time, necessitating efficient communication protocols. Additionally, the maximum transmission power is often capped at \( 14\,\mathrm{dBm} \) (\( 25\,\mathrm{mW} \)) equivalent isotropically radiated power (EIRP) to minimize interference with other spectrum users. To operate effectively within these regulatory constraints, LoRaWAN employs techniques like ADR, which dynamically adjusts EDs' data rate and transmission power based on network conditions \cite{erturkSurveyLoRaWANArchitecture2019}. This also optimizes network performance and capacity while ensuring compliance with regional spectrum policies.

\subsection{Classes of LoRaWAN Devices}

To accommodate various application requirements, LoRaWAN defines three classes of EDs, namely Class A, Class B, and Class C \cite{jagathaExploringLoRaWANClass2024}. Class A devices offer the lowest power consumption and are suitable for applications where downlink communication is infrequent or can be scheduled after an uplink transmission. After each uplink, a Class A device opens two short receive windows for potential downlink messages. In addition to Class A features, Class B devices open extra receive windows at scheduled times, synchronizing with the network server through periodic beacons. This lets the server know when the device is listening, which benefits applications requiring frequent downlink communication. Class C devices have almost continuous receive windows, closing only when transmitting. They are suitable for applications needing minimal latency for downlink communication but with higher power consumption.

\subsection{LoRaWAN in Indoor Environments}
\label{subsec: lorawan_indoor}

Implementing LoRaWAN in indoor settings presents unique challenges due to environmental factors that affect signal propagation. Obstacles such as walls, floors, and furniture can cause radio frequency (RF) signal attenuation, reflection, and multipath fading, weakening or distorting signals \cite{muppalaInvestigationIndoorLoRaWAN2021}. Building materials like concrete, glass, and metal may absorb or reflect radio signals, while thick walls and metallic structures such as reinforcements or ducts can increase path loss \cite{pavlikInfluenceVariousCommonly2024}. Additionally, dynamic factors, including human movement, the operation of electronic devices, and varying environmental conditions like relative humidity and temperature, can impact signal strength and quality \cite{keshmiryEffectsEnvironmentalOperational2023}. Therefore, understanding these factors is crucial for optimizing network deployment in indoor environments.

\section{Related Works}
\label{sec:related_work}

Research on LoRaWAN performance has predominantly focused on outdoor environments, utilizing PLMs to predict signal propagation. Theoretical models like the Friis Transmission Equation \cite{friisNoteSimpleTransmission1946} and the Log-Distance Path Loss Model (LDPLM) \cite{seidel914MHzPath1992} estimate signal attenuation based on distance and frequency but often neglect environmental factors such as obstacles and terrain variations. Empirical models like the Okumura-Hata model \cite{hataEmpiricalFormulaPropagation1980} incorporate measurement data to reflect actual signal behavior better, considering parameters like frequency, distance, antenna heights, and environmental correction factors \cite{kurtPathLossModelingWireless2017}. However, these models are less effective indoors, where signal propagation is influenced by additional factors not present outdoors.

Indoor environments present unique challenges due to multipath effects, wall attenuation, and diverse construction materials. Signals can reflect off walls and objects, be absorbed by materials like concrete and glass, and be shadowed by furniture and structural elements, significantly affecting signal strength \cite{azevedoCriticalReviewPropagation2024}. To address these complexities, specific indoor models have been developed. The LDPLM offers a simple approach for general predictions but may not adequately account for obstructions. The ITU-R P.1238 Model \cite{PropagationDataPrediction2021} estimates signal loss by incorporating operating frequencies and structural factors across rooms and walls. The Motley-Keenan Model \cite{limaMotleyKeenanModelAdjusted2005} accounts for multiple obstructions by considering the attenuation effects of walls and floors. Building upon this, the COST 231 Multi-Wall Model \cite{europeancommissionCOSTAction2311999} introduces wall-specific attenuation factors, providing detailed predictions in environments with diverse wall types. These models enhance the accuracy of signal propagation estimations in complex indoor spaces, facilitating effective LoRaWAN deployment.

\subsection{Previous Studies}

Over the past decade, extensive empirical and modeling studies have deepened our understanding of LPWAN technologies, particularly LoRaWAN's indoor behavior. Table~\ref{tab:relatedworks} summarizes key contributions, highlighting signal propagation characteristics, path loss modeling, and the impact of structural obstructions on performance metrics such as RSSI, SNR, SF, and PDR. This section synthesizes findings from real-world deployments and model-based analyses, offering a focused view of current capabilities and persistent challenges in indoor LoRaWAN environments.

The research in \cite{aksoyComparativeAnalysisEnd2024a} investigates the performance of LoRaWAN in a high-rise building, focusing on signal degradation across different floors. The study utilized EDs and an Adeunis Field Test Device (AFTD) to measure RSSI, SNR, and SF across four floors in an 18-floor building with the GW on the roof. Results revealed significant variation in signal quality, with RSSI and SNR values decreasing as distance from the GW increased. Higher SF values were frequently used on lower floors to maintain signal integrity. Physical obstructions, such as floors and walls, heavily influence LoRaWAN performance, and network optimization strategies are required to mitigate signal degradation in such environments.

Most recently, \cite{voAdvancePathLoss2024a} presents a dynamic PLM specifically tailored for LoRaWAN networks to improve distance estimation accuracy in indoor and urban environments. The indoor propagation analysis focused on RSSI measurements across different distances in a controlled \( 10 \, \mathrm{m} \times 8 \, \mathrm{m} \) room, showing that path loss increases with distance and environmental obstructions. For distances from \( x = 3 \, \mathrm{m} \) to \( x = 7 \, \mathrm{m} \), the RSSI data indicated significant fluctuations due to multipath fading, with path loss exponents varying accordingly. By employing an enhanced Kalman filter, the study reduced the noise in RSSI readings, achieving a mean error (ME) of \( 0.565 \, \mathrm{m} \) and a 90\% cumulative density function (CDF) error under \( 1.08 \, \mathrm{m} \). This precise modeling approach provides reliable distance estimations in indoor LoRaWAN setups.

In \cite{alkhazmiAnalysisRealWorldLoRaWAN2023a}, LoRaWAN performance in an eight-story residential building is evaluated. The study focused on non-line-of-sight (NLoS) communication, with a fixed receiver on the rooftop and the transmitter moved across different floors. The analysis revealed that RSSI and SNR improved significantly as the transmitter moved closer to the receiver on higher floors. Packet loss was negligible on most floors, demonstrating LoRaWAN's robustness in indoor environments. However, lower floors experienced some signal attenuation due to obstructions, though the overall network reliability remained strong.

In \cite{robles-encisoLoRaZigbee5G2023a}, the performance of LoRa, Zigbee, and 5G technologies in an indoor corridor under NLoS conditions was compared. LoRa showed strong resistance to multipath fading due to its CSS modulation. Close-In (CI) and Floating-Intercept (FI) models were used to analyze path loss, with LoRa performing better than Zigbee, particularly in more obstructed areas. The CI model showed LoRa with a path loss slope of 1.96 in NLoS-1 and 3.01 in NLoS-2. LoRa maintained stronger signal strength and higher packet delivery rates than Zigbee, making it more suitable for long-range indoor communications.

The experiments in \cite{harindaPerformanceLiveMultiGateway2022} evaluate the performance of LoRaWAN within a three-story office building in Glasgow. The indoor environment comprised open-plan floors with multiple GWs. Over 26 days, the results demonstrated a packet transfer success rate of 99.95\%, indicating excellent performance without significant interference. RSSI and SNR measurements showed consistent performance across different parts of the building, with minimal packet loss and mainly associated with obstacles like walls and doors. The study highlights the importance of GW placement in ensuring reliable coverage in indoor environments.

The study conducted by Muppala et al. \cite{muppalaInvestigationIndoorLoRaWAN2021} investigated LoRaWAN signal propagation in a city building, focusing on path loss due to indoor obstacles such as walls and doors. The research used RSSI and SNR measurements, showing that significant signal attenuation occurred, even over short distances. For instance, RSSI values dropped from \( -90\,\mathrm{dBm} \) to between \( -100\,\mathrm{dBm} \) and \( -120\,\mathrm{dBm} \) when moving from line-of-sight (LoS) locations to obstructed ones. The study also found anomalies in the signal due to multipath reflections at certain points, notably at distances around \( 40\,\mathrm{m} \), \( 80\,\mathrm{m} \), and \( 120\,\mathrm{m} \). These findings provide empirical evidence of the impact of physical obstructions on LoRaWAN performance in real-world indoor environments.

The limitations and performance of LoRaWAN in multi-story indoor environments are examined in \cite{sabanExperimentalAnalysisIoT2021a}. Tests at the University of Valencia revealed that while LoRaWAN could reliably transmit packets within short distances, its performance declined significantly as physical barriers such as walls and floors increased. Packet loss was substantial in locations farther from the GW, particularly beyond \( 70\,\mathrm{m} \), where more than 50\% of packets were lost. The study concludes that although LoRaWAN is effective for short-range indoor IoT applications, larger buildings or more obstructed environments require multiple GWs or optimized node placements to ensure reliable communication.

\begin{table*}[hbt!]
  \centering
  \caption{Recent Indoor LoRaWAN Path Loss Models: Variables, Methods, and Research Insights }
  \footnotesize
  \setlength{\tabcolsep}{4pt}
  \begin{tabularx}{\textwidth}{C{2.5em}C{2.5em}L{12em}L{18em}L{23.5em}}
    \toprule
    \textbf{Ref.} & \textbf{Year} & \textbf{Key Variables} & \textbf{Model/Approach} & \textbf{Contribution} \\
    \midrule
    \cite{aksoyComparativeAnalysisEnd2024a} & 2024 & RSSI, SNR, SF, distance, physical barriers (floors, walls) & Empirical comparison of performance metrics (RSSI, SNR, and SF) using both EDs and AFTD across different floors in an 18-floor building. & Delivered significant findings on LoRaWAN's performance under complex indoor conditions, demonstrating the effects of physical obstructions on signal strength and emphasizing the importance of optimized SF and ED placement to enhance coverage. \\
    \midrule
    \cite{voAdvancePathLoss2024a} & 2024 & RSSI, distance, path loss exponent (\(n\)), standard deviation, environmental noise (reflection, fading) & Dynamic PLM with enhanced Kalman filter to minimize RSSI noise for accurate distance estimation. & Achieved high-precision indoor path loss data for LoRaWAN using a dynamic model with an enhanced Kalman filter, resulting in low mean error and accurate distance estimation in noisy indoor environments. \\
    \midrule
    \cite{alkhazmiAnalysisRealWorldLoRaWAN2023a} & 2023 & RSSI, SNR, packet delivery, SF, physical obstructions (walls, floors) & Empirical analysis of LoRaWAN's indoor performance in an eight-story building under varying NLoS conditions. & Confirmed LoRaWAN's resilience in indoor multi-floor buildings with minimal packet loss, reinforcing its suitability for IoT applications in similar settings. \\
    \midrule
    \cite{robles-encisoLoRaZigbee5G2023a} & 2023 & RSSI, bit error rate, path loss, SF, physical obstructions (walls, corners) & Path loss and signal quality analysis using the CI and FI models, measuring power attenuation and performance for LoRa. & Showed LoRa's superior ability to maintain reliable communication in obstructed indoor environments, outperforming Zigbee and 5G over longer distances. \\
    \midrule
    \cite{harindaPerformanceLiveMultiGateway2022} & 2022 & RSSI, SNR, Packet success rate (PSR), Physical barriers (walls, doors), GW placement. & Real-world measurements of LoRaWAN performance in a live indoor environment using multiple GWs, focusing on PSR and interference levels. & Demonstrated that LoRaWAN can achieve high success rates indoors when GW placement is optimized to mitigate interference and improve coverage. \\
    \midrule
    \cite{sabanExperimentalAnalysisIoT2021a} & 2021 & RSSI, SNR, packet reception ratio (PRR), distance, physical obstacles (walls, floors) & An empirical evaluation of LoRaWAN's performance in a multi-floor building using real-world measurements of RSSI, SNR, and packet reception ratio across varying distances and floors. & Identified challenges in deploying LoRaWAN in complex indoor environments due to signal attenuation and packet loss and recommended optimized network architecture with strategic placement of GWs and nodes to maintain robust performance. \\
    \midrule
    \cite{muppalaInvestigationIndoorLoRaWAN2021} & 2021 & RSSI, SNR, distance, frequency, obstacles (walls, doors) & Empirical PLM derived from real-world indoor signal measurements & Provided indoor path loss data for LoRaWAN in complex building environments, highlighting the impact of obstacles on signal attenuation. \\
    \midrule
    \cite{elchallLoRaWANNetworkRadio2019a} & 2019 & RSSI, SNR, distance, floors, walls, antenna heights & Empirical PLM with a logarithmic increase of path loss and decreasing loss per floor & Proposed a highly accurate PLM for multi-floor indoor environments, showing reduced loss per floor and achieving 95\% PDR. \\
    \midrule
    \cite{muzammirPerformanceAnalysisLoRaWAN2019a} & 2019 & ToA, RSSI, packet loss, SFs, physical obstacles (walls, floors), communication range, network traffic & An empirical evaluation of LoRaWAN performance in a multi-floor indoor environment using real-world measurements of ToA, RSSI, and packet loss, with varying SF values and distances. & Analyzed LoRaWAN's indoor performance, showing over 90\% packet delivery despite signal attenuation from obstacles and network traffic. Recommended multi-gateway deployment to boost signal reliability indoors. \\
    \midrule
    \cite{erbatiAnalysisLoRaWANTechnology2018} & 2018 & RSSI, SNR, PDR, SF, distance, physical barriers (walls, ceilings) & Empirical analysis of LoRaWAN performance in an indoor multi-floor building, measuring RSSI, SNR, and PDR across varying distances and SFs. & Found LoRaWAN effective in multi-floor indoor environments but observed significant effects from physical barriers, especially in basements, on packet delivery and signal quality. \\
    \midrule
    \cite{ayelePerformanceAnalysisLoRa2017} & 2017 & RSSI, packet error rate (PER), SF, ToA, distance, physical obstacles (walls, NLoS locations) & Empirical performance evaluation of LoRa in an indoor setting, using EDs and a GW to measure RSSI, PER, and ToA across different SFs and distances. & Demonstrated LoRa's effectiveness in indoor settings but emphasized the necessity for careful SF configuration and strategic node placement to maintain performance in larger, complex buildings. \\
    \midrule
    \cite{hosseinzadehNeuralNetworkPropagation2017a} & 2017 & RSSI, path loss, attenuation, SF, physical barriers (walls, windows) & A hybrid propagation model using the COST231 model and an artificial neural network (ANN) to predict LoRaWAN signal behavior in indoor environments, validated with real-world measurements. & Showed that advanced modeling techniques, combining COST231 and ANN, are crucial for accurately predicting indoor LoRaWAN performance in obstructed environments, thereby improving network planning and deployment. \\
    \midrule
    \cite{petajajarviEvaluationLoRaLPWAN2017a} & 2017 & RSSI, PDR, SF, BW, physical barriers & Real-world testing of LoRaWAN in indoor environments using empirical measurements of RSSI and packet delivery across different SFs and BW settings. & Revealed LoRaWAN's suitability for indoor health monitoring applications while highlighting the critical importance of precise parameter configuration to mitigate signal attenuation due to structural obstacles. \\
    \bottomrule
    \end{tabularx}
  \setlength{\tabcolsep}{6pt}
  \label{tab:relatedworks}
\end{table*}

The work in \cite{elchallLoRaWANNetworkRadio2019a} comprehensively evaluates LoRaWAN network performance across urban, rural, and indoor environments in Lebanon, focusing on PLMs. In the indoor setting, the study measured signal attenuation in a multi-floor building, revealing that path loss increased logarithmically with distance but not linearly with the number of floors. For example, the loss per floor decreased from \( 8.1\,\mathrm{dB} \) on the second floor to \( 4.3\,\mathrm{dB} \) on the fourth floor. The proposed PLM, with an exponent of \(2.85\) and a reference path loss of \( 120.4\,\mathrm{dB} \), showed greater accuracy than traditional models like ITU-R and Cost 231-MWF. Additionally, the study reported an average PDR of 95\%, demonstrating LoRaWAN's effectiveness in deep indoor environments.

The research in \cite{muzammirPerformanceAnalysisLoRaWAN2019a} evaluates the performance of LoRaWAN in a multi-story indoor setting. The study, conducted at Universiti Teknologi MARA, involved placing LoRaWAN EDs across different floors of a building with a central GW to measure critical performance metrics such as ToA, RSSI, and packet loss. Results indicate that while LoRaWAN can maintain sufficient signal strength indoors, performance is significantly affected by physical obstructions like walls and floors. Increasing the SF led to higher ToA and packet loss, particularly in locations farther from the GW. The study concludes that LoRaWAN is suitable for IoT applications in multi-story environments but recommends using additional GWs to improve coverage and reliability.

LoRaWAN performance examinations in an indoor setting, focusing on the signal strength and PDR across various floors of a building, were carried out in \cite{erbatiAnalysisLoRaWANTechnology2018}. The indoor tests were conducted in a multi-story building with walls of gypsum and concrete ceilings. The study evaluated parameters such as RSSI and SNR across different SFs (SF7 to SF12). Results showed 100\% packet delivery at all measured points except in the basement, where PDR decreased, with the lowest being 62\% for SF7. Despite the basement signal challenges, the study demonstrated LoRaWAN's strong indoor performance in multi-floor environments, with minimal signal degradation on floors above ground.

In \cite{ayelePerformanceAnalysisLoRa2017}, an in-depth evaluation of LoRa radio performance in an indoor IoT environment is provided. The experiments were conducted in a multi-floor office building at the University of Twente. LoRa EDs were deployed across various distances from the GW to measure signal quality metrics like RSSI and PER. The study found that signal degradation increased with distance and physical obstructions, and packet error rates were higher in more distant NLoS locations. Higher SFs helped mitigate packet loss but also increased ToA, reducing throughput. The authors concluded that LoRa performs well for indoor IoT applications at shorter distances but requires optimized configurations for larger, more obstructed environments.

The design work in \cite{hosseinzadehNeuralNetworkPropagation2017a} evaluates LoRaWAN's signal propagation in an indoor environment at Glasgow Caledonian University. Measurements were collected on the seventh and eighth floors of a building, and a novel hybrid model combining the COST231 model with ANN was introduced to improve prediction accuracy. The study found that traditional models like log-distance and standard COST231 were insufficient in accurately predicting indoor signal attenuation. The hybrid model, which incorporated wall attenuation and LoS factors, significantly improved the prediction of RSSI values across various indoor locations. The study also highlighted that indoor propagation is heavily influenced by physical barriers such as walls and windows, necessitating careful model adjustments for indoor network planning.

Using LoRaWAN technology for indoor health monitoring applications was investigated in \cite{petajajarviEvaluationLoRaLPWAN2017a}. The study at the University of Oulu evaluated LoRaWAN performance across various indoor environments with concrete and steel structures. Using SF 12 and a \( 125\,\mathrm{kHz} \) BW, the results indicated a 96.7\% PDR across campus, even without acknowledgments or retransmissions. However, lower SFs led to greater packet loss, particularly at distances beyond \( 200\,\mathrm{m} \) from the base station. The research showed that while LoRaWAN can maintain indoor solid performance, its reliability diminishes as SF decreases and physical obstructions increase.

\subsection{Research Gap}

Despite growing interest in using LoRaWAN for indoor applications, comprehensive experimental campaigns are still lacking. Accurate prediction of indoor signal behavior remains challenging due to the highly variable nature of indoor spaces and overlooked dynamic environmental factors that affect signal attenuation by altering the medium's dielectric properties, as discussed in Section \ref{subsec: lorawan_indoor}. Most existing studies rely on limited datasets, transmitting only a few hundred to a thousand packets over short periods, with the longest campaigns lasting just 26 days. Additionally, these experiments are typically confined to specific locations and lack specialized models for office environments, failing to capture diverse propagation behaviors across varied indoor settings. This absence of extensive, long-term data and tailored modeling creates significant gaps in understanding LoRaWAN’s indoor performance and environmental dynamics. This highlights the need for larger-scale, more systematic investigations to assess LoRaWAN’s potential in such environments fully.

Also, in contrast, technologies like Wi-Fi and Zigbee have been extensively studied in indoor contexts, resulting in well-established models that account for indoor propagation characteristics \cite{zafariSurveyIndoorLocalization2019}. However, these technologies operate at higher frequencies and use different modulation schemes, leading to propagation behaviors that differ from LoRaWAN. Since LoRaWAN operates in sub-GHz frequency bands and employs CSS modulation, this interacts differently with indoor environments compared to higher-frequency technologies \cite{borLoRaLowPowerWideArea2016}. Consequently, models developed for Wi-Fi and Zigbee are not directly applicable to LoRaWAN.

These gaps highlight the necessity for empirical measurements tailored to specific indoor environments to develop accurate PLMs for LoRaWAN networks. To address this need, our study conducted a comprehensive six-month LoRaWAN measurement campaign within an indoor office setting during the mid-fall and winter seasons. We deployed one GW and six fixed EDs within a 40-meter range, utilizing various wall types while maintaining similar antenna heights to emulate typical office layouts.

We collected \( 1,\!328,\!334 \) data fields, capturing not only LoRaWAN transmission metadata but also factors such as frequency, SF, and signal propagation metrics, including RSSI, SNR, and ToA. Additionally, we calculated metrics such as ESP and NP based on the RSSI and SNR features. Our dataset also includes design parameters like distance, as well as environmental conditions including temperature, relative humidity, barometric pressure, CO\textsubscript{2} concentration, and (PM\textsubscript{2.5}). This data descriptor includes parameters of ED and GW antenna heights (above the floor), transmitter power, antenna gains, and frame lengths, which are crucial for data analysis and modeling.

Leveraging this extensive dataset, we proposed and validated an enhanced PLM based on the COST231 Multi-Wall Model \cite{europeancommissionCOSTAction2311999}, demonstrating one of the ways how this dataset can be utilized. By integrating these environmental factors, our model provides a more precise and accurate understanding of indoor signal propagation, significantly improving the reliability of LoRaWAN deployments in complex office environments.

\section{Experimental Design Methodology}
\label{sec: methodology}

\begin{figure}[hbt!]
\centering
\centerline{\includegraphics[width=\columnwidth]{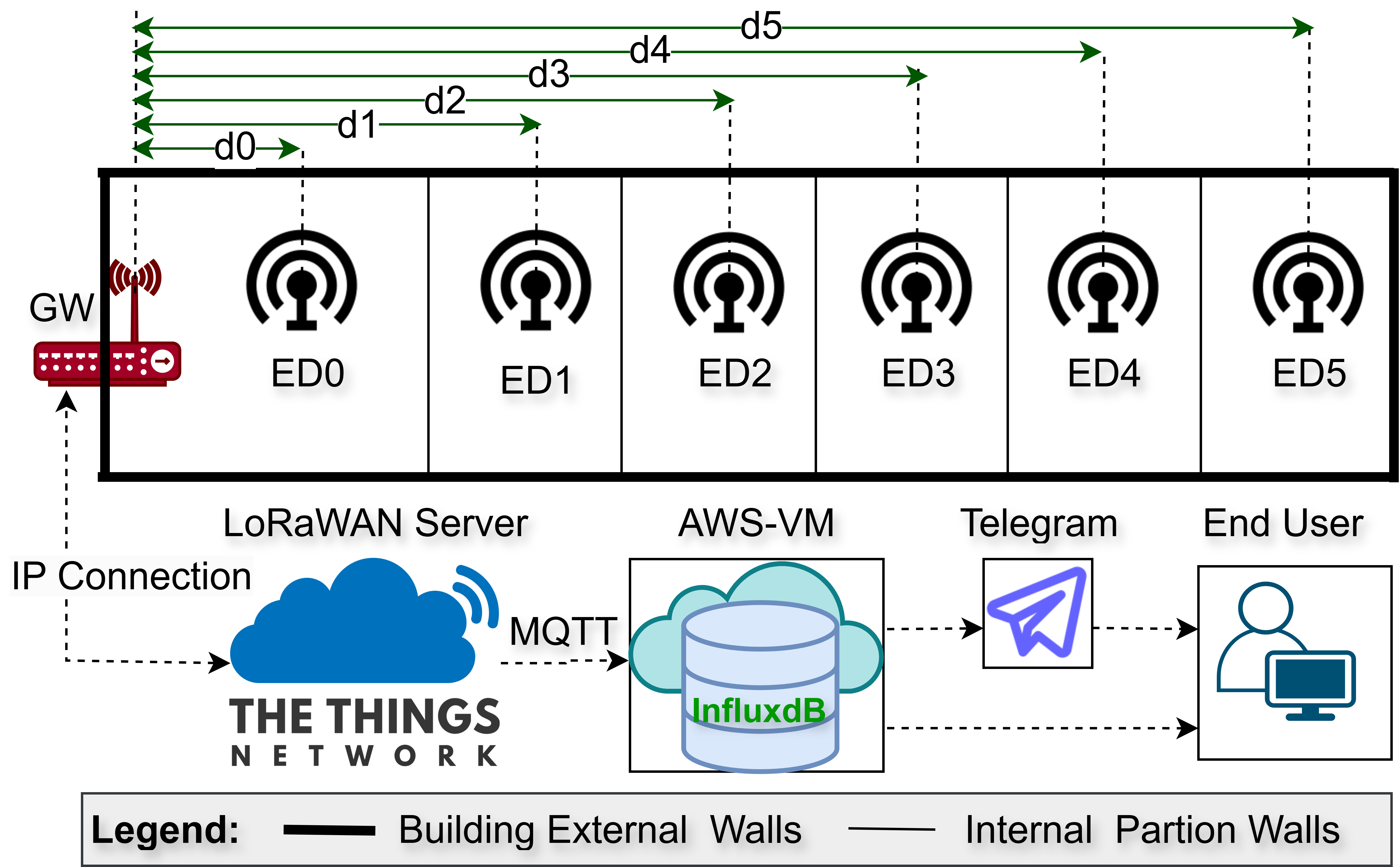}}
\caption{\label{fig: design} {Experimental End Device (ED) and Gateway (GW) Deployment Layout (Not drawn to Scale): EDs $ \text{ED0--ED5} $ and the GW are placed in an indoor environment with brick/concrete and wooden walls. The GW connects to The Things Network (TTN), sending data via a Message Queuing Telemetry Transport (MQTT) broker to InfluxDB on an Amazon Web Services (AWS) virtual machine (VM), with data logging failure alerts sent via Telegram.}}
\end{figure}

This study conducted a series of systematic measurements to assess LoRaWAN path loss in an indoor office setting. The experimental setup included strategically placed LoRaWAN EDs and the GW as depicted in Fig.~\ref{fig: design}. Data were collected under different conditions to simulate varying levels of environmental changes. This section details the equipment, experimental design, and data collection procedures.

\begin{figure*}[hbt!]
\centering
\centerline{\includegraphics[width=\textwidth]{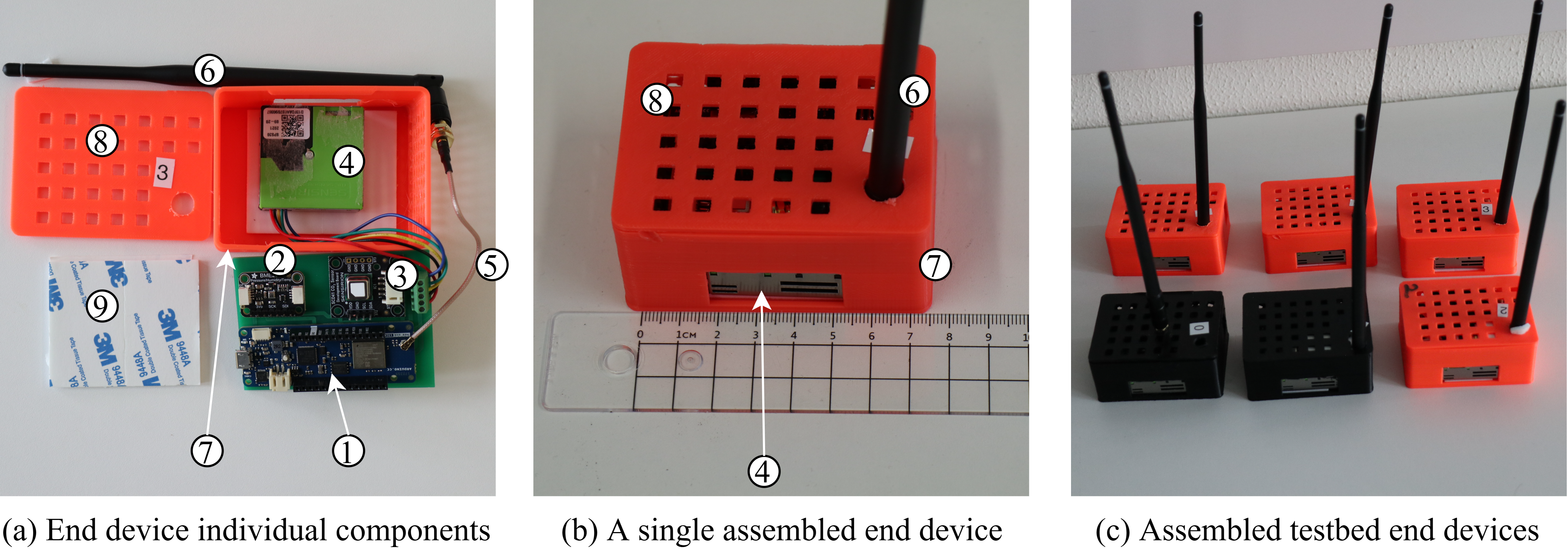}}
\caption{\label{fig: end_device} {End Devices used in the Campaign; Components and Assembly: (1) Arduino MKR WAN 1310, (2) Adafruit BME280 sensor, (3) Sensirion SCD41 sensor, (4) Sensirion SPS30 sensor, (5) SMA to uFL adapter cable, (6) Rubber Duck antenna, (7) 3D-printed casing base, (8) 3D-printed casing lid, (9) mounting adhesive pads.}}
\end{figure*}

\subsection{Hardware Description}

\subsubsection{End Devices}

 A sample of the ED is shown in Fig.~\ref{fig: end_device}. We designed a custom 3D-printed enclosure in our hardware lab, incorporating perforations to promote airflow and mitigate heat buildup for more accurate sensor measurements. These vents also help minimize unwanted signal reflections while ensuring secure housing for the electronics and antenna. The lightweight and sturdy structure maintains mechanical protection and thermal stability, making it well-suited for our deployment.
 
 For this campaign, we utilized the Arduino MKR WAN 1310 platform due to its robust processing and communication capabilities as the ED microcontroller unit. It integrates a SAMD21G18A processor, featuring an ARM Cortex M0+ running at up to \( 48\,\mathrm{MHz} \), complemented by \( 256\,\mathrm{kB} \) of flash memory and \( 32\,\mathrm{kB} \) of SRAM. Central to its communication functionality is the Murata CMWX1ZZABZ LoRa module, which includes an STM32L0 series ARM Cortex M0+ microcontroller and a Semtech SX1276 radio, delivering up to \( +20\,\mathrm{dBm} \) RF  output power (\( \mathrm{TP}_{\mathrm{TX}} \)). The LoRaWAN stack is configured to operate in Class A mode, optimizing energy efficiency and enabling bidirectional communication with minimal latency. For our application, we opted for a lower power usage setting of \( 20\,\mathrm{dBm} \) to balance performance and energy efficiency and adhere to regional regulations. This module supports multiple frequency bands of \(433\), \(868\), and \( 915\,\mathrm{MHz} \), making it adaptable to various regional requirements. Additionally, it incorporates a cryptographic co-processor (ATECC508) for secure hardware-based key storage and an extra \( 16\,\mathrm{Mbit} \) Serial Peripheral Interface (SPI) flash memory for extended storage capabilities. Its compatibility with various interfaces, including Inter-Integrated Circuit (I2C), SPI, and Universal Asynchronous Receiver/Transmitter (UART), enhances its versatility for our environmental sensor network applications.

Each of the six identical EDs was equipped with a protruding omnidirectional antenna with a peak gain ($G_{TX}$) of approximately \( 0.4\,\mathrm{dBi} \), designed to minimize internal reflections and reduce signal distortion in line with the development board's strict specifications. These antennas are constructed from robust materials and feature SMA or RP‐SMA connectors that facilitate easy interconnection with the boards. However, the effective antenna gain is slightly reduced by the insertion loss of the \( 10\,\mathrm{cm} \) cable connecting to the uFL port. In our setup, the TUOLNK RP‐SMA to uFL cable employed an RG178 coaxial cable, which typically exhibits an attenuation (\( \mathrm{CL}_{\mathrm{TX}} \)) of about \(1.4\,\mathrm{dB/m}\) at \( 868\,\mathrm{MHz} \), translating to roughly \(0.14\,\mathrm{dB}\) loss over a \( 10\,\mathrm{cm} \) length. The antennas were mounted perpendicular to the ground using support lines to maintain stability and ensure optimal performance.

\subsubsection{Environmental Sensors}

Three advanced sensors were integrated into MKR WAN 1310 platform to measure temperature, relative humidity, CO\textsubscript{2}, barometric pressure, and PM\textsubscript{2.5} levels. We selected the I2C interface for all sensors to leverage their onboard analog-to-digital conversion and shared bus architecture. This approach significantly reduces wiring complexity, allowing multiple devices to communicate on the same lines, making it ideal for precise, scalable indoor environmental monitoring. The design and operation specifications of the sensors deployed are briefly described below.

\begin{itemize}
    \item \textbf{An SCD41 Sensor}: Developed by Sensirion, measures temperature, relative humidity, and CO\textsubscript{2} levels using photoacoustic sensing technology. It can measure CO\textsubscript{2} concentrations ranging from \( 400 \) to \( 5000\,\mathrm{ppm} \) with an accuracy of \(\pm (50\,\mathrm{ppm} + 5\%\text{ of reading})\). The embedded temperature and relative humidity sensor provide measurements within the ranges of \(-10\)\textdegree{}C to \( 60\)\textdegree{}C for temperature with an accuracy of \(\pm 1.5\)\textdegree{}C, and \( 0\%\) to \( 100\%\) for relative humidity with an accuracy of \(\pm 9\%\). The sensor's compact size (\( 10.1\,\mathrm{mm} \times 10.1\,\mathrm{mm} \times 6.5\,\mathrm{mm} \)) and surface-mount device packaging allow for easy integration into a breakout board of dimensions \( 28\,\mathrm{mm} \times 24\,\mathrm{mm} \times 8\,\mathrm{mm} \). It supports I2C and UART communication interfaces, low-power operation modes, and features such as automatic self-calibration, ensuring long-term stability and minimal maintenance.
    \item \textbf{An Adafruit BME280 Sensor}: Based on Bosch's advanced environmental sensor technology, provides precise barometric pressure measurements with an absolute accuracy of \(\pm 1.0\,\mathrm{hPa}\) and operates within a barometric pressure range of \( 300\,\mathrm{hPa} \) to \( 1100\,\mathrm{hPa} \). It supports both I2C and SPI communication protocols, facilitating easy integration with sensors and microcontrollers. The compact design (\( 2.5\,\mathrm{mm} \times 2.5\,\mathrm{mm} \times 0.93\,\mathrm{mm} \)) and high precision make it an excellent choice for detailed environmental monitoring in indoor office settings.
    \item \textbf{A Sensirion SPS30 Sensor}: This state-of-the-art particulate matter sensor measures PM\textsubscript{2.5} levels using laser scattering and innovative contamination-resistance technology, ensuring precise and reliable measurements over a prolonged lifespan exceeding ten years. It detects PM\textsubscript{2.5} particles within a size range of \( 0.3 \) to \( 2.5\,\mu\mathrm{m} \) and provides mass concentration measurements from \( 0 \) to \( 100\,\mu\mathrm{g}/\mathrm{m}^3 \) with an accuracy of \(\pm 10\%\) for concentrations between \( 100 \) to \( 1000\,\mu\mathrm{g}/\mathrm{m}^3 \). Its compact dimensions (\( 41\,\mathrm{mm} \times 41\,\mathrm{mm} \times 12\,\mathrm{mm} \)) make it exceptionally space-efficient. It supports I2C and UART communication interfaces, facilitating easy integration into various environmental monitoring systems. We configured the sensor for self-fan cleaning every four days, ensuring consistent performance and longevity.
\end{itemize}

\subsubsection{The Gateway}

\begin{figure}[hbt!]
\centering
\centerline{\includegraphics[width=0.7\columnwidth]{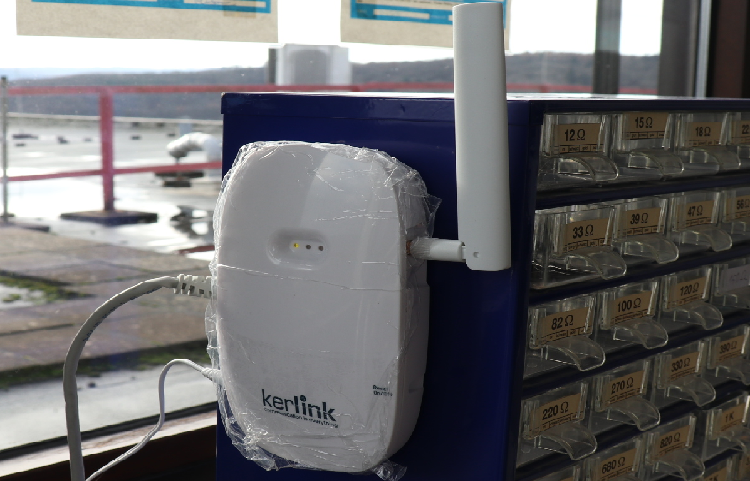}}
\caption{\label{fig: gw} {Wirnet iFemtoCell Gateway (GW) used for LoRaWAN connectivity during our data campaign.}}
\end{figure}

The Wirnet iFemtoCell, a robust indoor LoRaWAN gateway configured for the \(868\,\mathrm{MHz}\) band, was employed in this study due to its suitability for European IoT deployments. As shown in Fig.~\ref{fig: gw}, the physical setup demonstrates its deployment during our experiments. We programmed it with KerOS 4.3.3, leveraging its advanced features including \(49\) LoRa demodulators over \(9\) channels, operating within the \( 863{-}873\,\mathrm{MHz} \) band, and achieving a sensitivity down to \(-141\,\mathrm{dBm}\) at SF12. The GW uses a \( 868/915\,\mathrm{MHz} \) Rubber Dipole omnidirectional antenna with \( 3\,\mathrm{dBi} \) gain ($G_{RX}$) and vertical polarization, connected directly via an RP-SMA connector, ensuring efficient signal transmission and reception. Therefore, the receiver cable losses (\( \mathrm{CL}_{\mathrm{RX}} \)) were neglected. It supports Ethernet for back-haul connectivity and offers high performance in demanding environments such as basements and underground parking. With its compact, durable design, built-in high-rejection filters, and remote management capabilities via the Wanesy Management Center, the Wirnet iFemtoCell is a reliable choice for our IoT communication needs. We used the LoRa packet forwarder and registered the GW to a readily configured The Things Network (TTN) LoRaWAN server.

\begin{table*}[hbt!]
\centering
\caption{Frame Configuration Sent from end devices (EDs) to the gateway (GW)}
\label{tab:frame_configuration}
\begin{tabular}{ccccccc}
\toprule
\textbf{Variable} & \textbf{Data Type} & \textbf{Max. Value} & \textbf{Max. Value x 100} & \textbf{Bit Representation} & \textbf{\# Bits} \\
\midrule
Temperature (\textdegree C) & int16\_t       & 100           & 10000    & 10011100010000    & 14        \\
Relative Humidity (\%)     & int16\_t       & 100   & 10000     & 10011100010000          & 14         \\
Barometric Pressure (hPa)   & int32\_t       & 1100    & 110000    & 110101100100000      & 17         \\
CO\textsubscript{2} (ppm)        & uint16\_t      & 5000           & -       & 1001110001000      & 13     \\
PM$_{2.5}$ (\textmu g/m\textsuperscript{3})    & int16\_t   & 1000  & 100000 & 11000011010100000    & 17  \\
Packet Count     & uint32\_t      & 4294967295  & -       & $2^{32} - 1$ & 32  \\
\midrule
\textbf{Total}        &               &                &                     &                           & \textbf{107}    \\
\bottomrule
\end{tabular}
\end{table*}

\subsection{Software Design Description}

\subsubsection{LoRaWAN Frame Configuration}

Each ED was configured to package sensor data into a binary frame format in Table~\ref{tab:frame_configuration} for efficient LoRa transmission. The main variables in the LoRaWAN frame included temperature, relative humidity, barometric pressure, CO\textsubscript{2}, PM$_{2.5}$, and a packet count. Temperature, relative humidity, barometric pressure, and PM$_{2.5}$ values were scaled by \(100\) to maintain one or two decimal places of precision. The variables were also represented using 16-bit integers (\texttt{int16\_t}) for temperature, relative humidity, and PM$_{2.5}$, while barometric pressure used a 32-bit integer (\texttt{int32\_t}). The CO\textsubscript{2} value was directly included as a 16-bit integer (\texttt{uint16\_t}) without scaling since it was already in a suitable range and precision. The packet-count is a 32-bit unsigned integer (\texttt{uint32\_t}), ensuring robust sequence tracking whose maximum value could not easily be predetermined. The payload was constructed by splitting each 16-bit and 32-bit value into high and low bytes to fit the transmission format, ensuring that all data is accurately represented within the allocated \( 18\,\text{bytes} \). The actual data, however, totals \( 107\,\text{bits} \) (\( 14\,\text{bytes} \)), and the remaining bits are reserved for future use or as padding, resulting in a total packet size of \( 18\,\text{bytes} \) (\( 144\,\text{bits} \)). This approach ensures efficient use of the payload space while allowing for future data expansion, which is recommended in LoRaWAN to maintain flexibility and adaptability in dynamic environments, making it ideal for our campaign.

\subsubsection{Transmission Protocol and Duty Cycle Compliance}
To guarantee fair use of SFs, we disabled LoRaWAN’s ADR and assigned the SF values manually. This prevents ADR-induced SF changes, enabling unbiased performance analysis under varying link conditions. In compliance with EU868 ISM regional parameters, we ensured our transmission protocol maintains the \( 1\% \) duty cycle in the \( 868.0 \) to \( 868.6\,\mathrm{MHz} \) range, with each device sending an 18-byte payload every \( 60\,\mathrm{seconds} \) as follows: Equation (\ref{eq: toa}) is used to determine \(ToA\), as per \cite{semtechcorporationSX127677782020}.
\begin{equation}
\label{eq: toa}
    \mathrm{ToA} = \left( N_{\mathrm{preamble}} + 4.25 \right) \times T_{\mathrm{symbol}} + N_{\mathrm{PL}} \times T_{\mathrm{symbol}} ,
\end{equation}
where \( N_{\text{preamble}} \) represents the number of preamble symbols and \( N_{\text{PL}} \) denotes the number of payload symbols calculated based on specific transmission parameters using  (\ref{eq: n_payload}) as follows:
{\footnotesize
\begin{align}
N_{\mathrm{PL}} &= 8 + \max\left(\left\lceil \frac{8 \times \mathrm{PL} - 4 \times \mathrm{SF} + 28 + 16 \times \mathrm{CRC} - 20 \times H}{4 \times (\mathrm{SF} - 2 \times \mathrm{DE})} \right\rceil \right. \nonumber \\
&\quad \left. \times (\mathrm{CR} + 4), 0\right) ,
\label{eq: n_payload}
\end{align}}
where \( \mathrm{PL} \) refers to the payload size in bytes, varying from \(1\) to \(255\). \( \mathrm{SF} \) ranges from \(6\) to \(12\). \( \mathrm{CRC} \) stands for Cyclic Redundancy Check, a two-byte field that can optionally be appended to the LoRa payload to detect errors. If \( \mathrm{CRC} = 1 \), a 16-bit CRC is appended to the payload; otherwise, if \( \mathrm{CRC} = 0 \), no CRC is added. The \( H \) parameter indicates the header mode, where \( H = 0 \) signifies an explicit header and \( H = 1 \) indicates an implicit header. The \( \mathrm{DE} \) stands for Low Data Rate Optimization, which is enabled when \( \mathrm{DE} = 1 \) and disabled when \( \mathrm{DE} = 0 \). Lastly, \( \mathrm{CR} = 1 \) corresponds to \( 4/5 \) and \( \mathrm{CR} = 2 \) corresponds to \( 4/6 \).

As an example, for \( \mathrm{SF} = 7 \) and a \( \mathrm{BW} \) of \( 125\,\mathrm{kHz} \), using~(\ref{eq: symbolduration}), we found that \( T_{\mathrm{symbol}} = 1.024\,\text{ms} \). Also, for our case where \( \mathrm{PL} = 18 \), \( \mathrm{CRC} = 1 \), \( H = 1 \), \( \mathrm{DE} = 0 \), and \( \mathrm{CR} = 1 \), \( N_{\mathrm{PL}} \) is calculated to be \( 33\,\text{symbols} \). Using~(\ref{eq: toa}), we determined the \( \mathrm{ToA} \) to be \( 46.336\,\text{ms} \). Therefore, the total ToA amounts to \( 231.68\,\text{ms per hour} \) for five transmissions.

 Overall, for our configuration, we determined that five transmissions per hour across SF7 to SF12 at a \( 125\,\mathrm{kHz} \) BW result in a total ToA of approximately \( 18{,}302.5\,\mathrm{ms} \) per hour, equating to a duty cycle of about \( 0.503\% \), well below the regulatory limit. When scaling to six devices, our 8-channel Kerlink iFemto GW effectively distributes the combined airtime across all channels, maintaining each channel's duty cycle at around \( 0.377\% \). This assumes that transmissions are evenly spread and interference remains minimal, ensuring each channel stays comfortably within the duty cycle threshold as our network grows. Our approach ensures compliance with EU868 regulations while demonstrating scalability and reliability for robust indoor LoRaWAN communication.

\subsubsection{Cloud Storage and Data Management}

To manage and store LoRaWAN field measurements, we established and implemented a comprehensive cloud-based data management pipeline. Initially, a custom JavaScript formatter was developed and configured on the TTN server to decode the payloads sent by each ED. The server then exposes this decoded data and publishes it through a message queuing telemetry transport (MQTT) broker. To capture these published messages in real time, we employed a Python-based MQTT subscriber script, which was securely authenticated using an API key. The MQTT protocol was specifically chosen for its robustness in low-BW, high-latency environments, which is ideal for IoT applications like LoRaWAN. This pipeline was deployed on an AWS elastic compute cloud (EC2) instance, specifically a \texttt{t3.micro} VM configured with 2 vCPUs. The \texttt{t3.micro} instance was selected for its balance of cost-effectiveness and sufficient computational power for our data processing needs. The VM also had an InfluxDB v2.7.8 client for Ubuntu, configured to handle the time-series data storage requirements, leveraging InfluxDB's optimized performance for handling large volumes of time-stamped data.

Throughout the measurement campaign, this pipeline ran continuously on the AWS Elastic Compute Cloud instance, ensuring uninterrupted data flow. Access to the VM for setup and maintenance was secured via a secure shell key, with operations conducted through the VM's IP address. Using InfluxQL, with the Structured Query Language-like syntax, initial data querying was performed directly on the InfluxDB instance to validate and inspect the incoming measurements. A Python script was configured to query the database and export data into CSV files for advanced exploratory data analysis (EDA) and modeling. This method provided enhanced flexibility for data manipulation, integration with Python libraries, and automated processing, making it more suitable for handling a large dataset.

\begin{figure}[hbt!]
\centering
\centerline{\includegraphics[width=0.6\columnwidth]{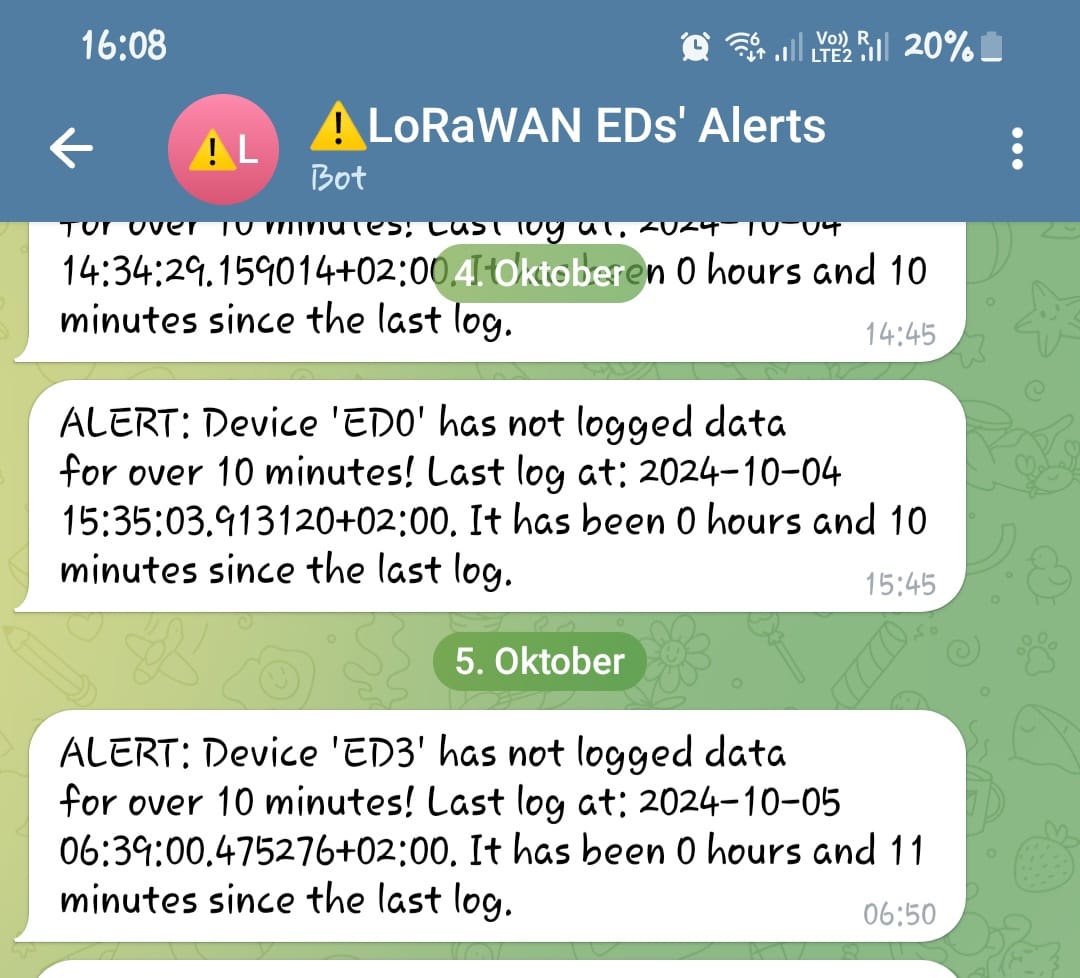}}
\caption{\label{fig:alerts} {Automated End Devices Data Gap Alerts: Telegram bot notifications example for data logging gaps over 10 minutes.}}
\end{figure}

Additionally, we configured a data-logging alert pipeline to monitor the EDs. Since InfluxDB was employed for time-series data storage, a Python script accessing this database was developed to detect data gaps. Running every five minutes on our AWS VM, the script generated alerts via a Telegram bot with secure token-based authentication whenever ED data was missing for more than ten minutes (see Fig.~\ref{fig:alerts}). This setup allowed us to monitor events such as ED power-offs, network outages, and AWS VM reachability issues. Consequently, it facilitated troubleshooting and ensured a reliable data collection campaign.

\subsection{Sensor Network Deployment}

\begin{figure}[hbt!]
\centering
\centerline{\includegraphics[width=\columnwidth]{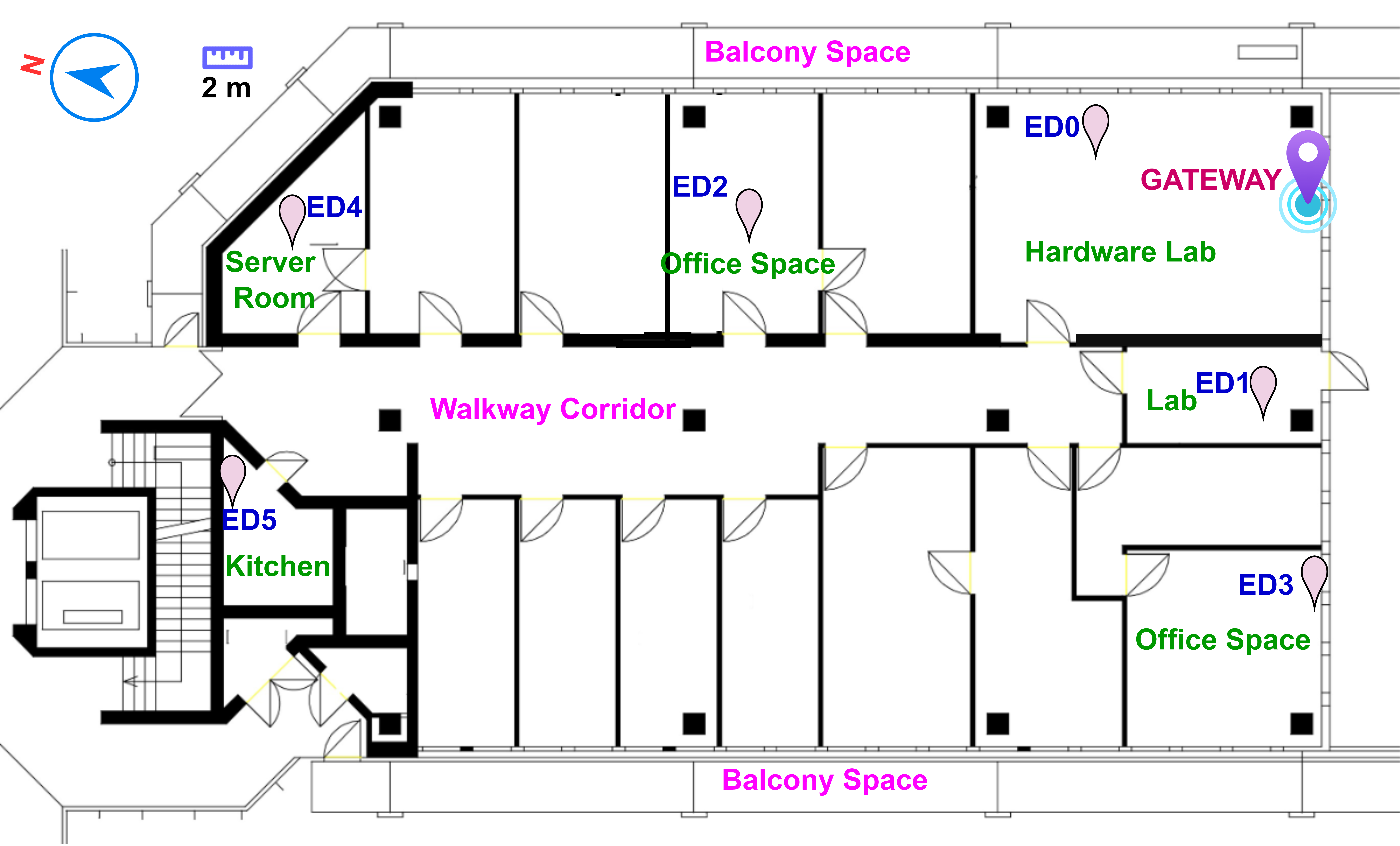}}
\caption{\label{fig:snd} Sensor Network Deployment: A layout showing sensor nodes (ED0–ED5) and the gateway (GW) placement in the indoor environment.}
\end{figure}

The deployment of six EDs (ED0 to ED5) is situated on the 8th floor of a building at the Hölderlinstraße Campus, University of Siegen in the City of Siegen, Germany, as shown in Fig.~\ref{fig:snd}. The floor plan follows a typical modern rectangular office layout, featuring corridors and partitions that create LoS and NLoS pathways. All devices are uniformly installed at approximately \( 0.8\,\mathrm{m} \) above the floor level (\( h_{TX} \)). Their placement relative to the GW varies, with distances ranging from \( 8{-}40\,\mathrm{m} \). Table~\ref{tab:network_deployment} provides detailed information on each device's location, distance to the GW, and the types of wall obstructions present.

\begin{table}[hbt!]
  \centering
  \caption{Network Deployment Details of End Devices (EDs) relative to the Gateway (GW)}
    \begin{tabular}{>{\centering\arraybackslash}m{1.2cm}>{\centering\arraybackslash}m{1.8cm}>{\centering\arraybackslash}m{1.5cm}>{\centering\arraybackslash}m{2cm}} 
    \hline
    \multirow{2}{*}{\textbf{Device}} & \multirow{2}{*}{\textbf{Distance (m)}} & \multicolumn{2}{c}{\textbf{Type of Wall}} \\
    \cline{3-4}
          &          & \textbf{Brick Wall} & \textbf{Wood Partitions} \\
    \hline
    ED0 & 10 & 0 & 0 \\
    ED1 & 8  & 1 & 0 \\
    ED2 & 25 & 0 & 2 \\
    ED3 & 18 & 1 & 2 \\
    ED4 & 37 & 0 & 5 \\
    ED5 & 40 & 2 & 2 \\
    \hline
    \end{tabular}
  \label{tab:network_deployment}
\end{table}

Deploying a LoRaWAN network in an indoor environment presents unique challenges due to obstacles that impede wireless signal propagation. Primary obstructions in such deployments include thin concrete and brick walls and wood partitions, which cause varying levels of signal attenuation and significantly impact overall network performance \cite{azevedoCriticalReviewPropagation2024}. Previous studies, analyzed in Table~\ref{tab:relatedworks}, have demonstrated effective strategies for deploying LoRaWAN networks in similar indoor settings. These studies address signal attenuation and multipath effects issues, providing valuable insights into our optimizing network design and device placement. Building on these findings, we chose the EDs and GW link budget parameter allocations as summarized in Table~\ref{tab:link_budget}, incorporating critical factors such as cable losses and antenna gains to ensure reliable coverage despite the challenging indoor environment.

\begin{table}[hbt!]
    \centering
    \caption{Link Budget Parameter Allocations for End Devices and the Gateway}
    \begin{tabular}{ccc}
    \hline
        \textbf{Parameter} & \textbf{Unit} & \textbf{Value} \\
    \hline  
        End device transmit power (\( \mathrm{TP}_{\mathrm{TX}} \)) & dBm & \(14\) \\
        End device cable losses (\( \mathrm{CL}_{\mathrm{TX}} \)) & dB & \(0.14\) \\
        End device peak antenna gain $(G_{TX})$ & dBi & \(0.4\) \\
        End device antenna height (above the floor) $(h_{TX})$ & m & \(0.8\) \\
        Gateway cable losses (\( \mathrm{CL}_{\mathrm{RX}} \)) & dB & \(0\) \\
        Gateway peak antenna gain $(G_{RX})$ & dBi & \(3\) \\
        Gateway antenna height (above the floor) $(h_{RX})$ & m & \(1\) \\
    \hline    
    \end{tabular}
    \label{tab:link_budget}
\end{table}

The strategic placement of devices was carefully designed to balance coverage and reliability. ED0 and ED1, positioned closest to the GW at distances of \( 10\,\mathrm{m} \) and \( 8\,\mathrm{m} \), respectively, were chosen for their minimal obstructions. ED0 has a direct line of sight to the GW, while ED1 is separated by a single thin concrete and brick wall, which may introduce slight signal attenuation. ED2 and ED3 are placed \( 25\,\mathrm{m} \) and \( 18\,\mathrm{m} \) from the GW at intermediate distances, respectively. ED2 faces two wood partitions, whereas ED3 is obstructed by one thin concrete and brick wall along with two wood partitions, providing an opportunity to evaluate network performance in environments with moderate signal barriers. ED4 and ED5, located farthest from the GW at \( 37\,\mathrm{m} \) and \( 40\,\mathrm{m} \), are subjected to the most challenging conditions. Five wood partitions separate ED4, and ED5 encounters two thin concrete and brick walls and two wood partitions. This placement is intended to assess the network's robustness under heavy obstruction scenarios with varied material types.

Several factors were considered while deploying the LoRaWAN network indoors to ensure optimal performance. Signal attenuation is a key factor, typically represented as path loss (\( \mathrm{PL} \)). Different materials cause varying degrees of path loss, where thin concrete and brick walls (\( \mathrm{PL}_{\mathrm{concrete}} \)) are expected to attenuate signals more significantly than wood partitions (\( \mathrm{PL}_{\mathrm{wood}} \)). Understanding these effects aids in predicting coverage areas and identifying potential dead zones. Additionally, multipath effects (\( \mathrm{PL}_{\mathrm{multipath}} \)) arise due to reflections and scattering caused by multiple obstructions, potentially degrading the quality and reliability of communication links. Finally, environmental parameters such as relative humidity (\( \mathrm{RH} \)),  temperature (\( T \)), and CO\textsubscript{2}, though generally stable indoors, can still influence signal propagation by induction in relation to occupancy effects. Monitoring these parameters ensures a comprehensive understanding of network performance and helps maintain reliable communication across the deployment area.

\begin{table*}[hbt!]
\centering
\caption{Descriptive statistics of environmental and signal quality dataset fields for LoRaWAN indoor propagation analysis}
\label{tab:stat_description}
\begin{tabular}{ccccccccc}
\toprule
\textbf{Field} & \textbf{Unit} & \textbf{Mean} & \textbf{STD ($\sigma$)} & \textbf{Min} & \textbf{1\textsuperscript{st} Quartile} & \textbf{Median} & \textbf{3\textsuperscript{rd} Quartile} & \textbf{Max} \\
\midrule
co2 & ppm & 553.934 & 136.080 & 386 & 449 & 513 & 625 & 1993 \\
humidity & \% & 37.544 & 6.493 & 13.99 & 32.81 & 37.81 & 41.51 & 60.19 \\
pm25 & µg/m\textsuperscript{3} & 1.982 & 2.542 & 0.0 & 0.63 & 1.27 & 2.41 & 637.71 \\
pressure & hPa & 323.321 & 10.805 & 286.91 & 316.54 & 324.33 & 331.49 & 347.57 \\
temperature & $^\circ$C & 21.207 & 2.580 & 13.74 & 19.73 & 21.17 & 22.53 & 40.24 \\
rssi & dBm & -76.903 & 22.899 & -128 & -93 & -73 & -60 & -28 \\
snr & dB & 7.419 & 7.098 & -24.5 & 7.8 & 9.5 & 11.2 & 19.0 \\
SF & bit/sym & 9.323 & 1.686 & 7 & 8 & 9 & 11 & 12 \\
frequency & MHz & 867.830 & 0.462 & 867.1 & 867.5 & 867.9 & 868.3 & 868.5 \\
toa & s & 0.556 & 0.589 & 0.072 & 0.134 & 0.247 & 0.987 & 1.974 \\
distance & m & 22.746 & 12.294 & 8 & 10 & 23 & 36 & 40 \\
c\_walls & - & 0.674 & 0.751 & 0 & 0 & 1 & 1 & 2 \\
w\_walls & - & 1.828 & 1.664 & 0 & 0 & 2 & 2 & 5 \\
exp\_pl & dB & 94.163 & 22.899 & 45.26 & 77.26 & 90.26 & 110.26 & 145.26 \\
n\_power & dBm & -86.025 & 20.472 & -141.212 & -102.738 & -84.224 & -70.467 & -33.146 \\
esp & dBm & -78.605 & 25.445 & -146.515 & -93.877 & -73.574 & -60.414 & -28.396 \\
\bottomrule
\end{tabular}
\end{table*}

\section{Exploratory Data Analysis}
\label{sec: data_description}

The following list provides a detailed description of each column in the dataset used for LoRaWAN path loss measurements in an indoor office setting, including the corresponding units where applicable. The dataset comprises \( 1,\!328,\!334 \) data fields, offering a high-resolution foundation for statistical inference under varied indoor conditions.

\begin{enumerate}
    \item \textbf{time}: Timestamp indicating when the current observation was recorded, units: yyyy-mm-dd hh:mm:ss (timezone: Europe/Berlin).
    
    \item \textbf{device\_id}: Unique identifier for the ED involved in the current measurement.
    
    \item \textbf{co2}: Measured concentration of (CO\textsubscript{2}) in the environment in parts per million (ppm).
    
    \item \textbf{humidity}: Relative humidity of the environment, indicating the amount of moisture in the air in \%.
    
    \item \textbf{pm25}: Concentration of particulate matter with a diameter of less than \( 2.5\,\mu\mathrm{m} \) in the environment in micrograms per cubic meter ($\mu$g/m$^3$).
    
    \item \textbf{pressure}: Barometric pressure of the environment in hectopascals (hPa).
    
    \item \textbf{temperature}: Ambient temperature of the environment where the measurement was taken in $^\circ$C.
    
    \item \textbf{rssi}: RSSI at the GW, representing the power level of the received signal in dB relative to one milliwatt (dBm).
    
    \item \textbf{snr}: SNR, which compares the desired signal level to the level of background noise in dB.
    
    \item \textbf{SF}: SF used in the transmission, determining the duration of each data symbol and thus the data rate in bits/symbol.
    
    \item \textbf{frequency}: Carrier frequency at which the signal was transmitted in Megahertz (MHz).
    
    \item \textbf{f\_count}: Frame counter used to confirm transmissions for  PDR calculation.
    
    \item \textbf{p\_count}: Count of packets transmitted during the measurement period for each ED.
    
    \item \textbf{toa}: Time on Air (ToA), the duration for which the transmission occupied the channel in seconds (s).
    
    \item \textbf{distance}: Distance between the GW and the corresponding ED in meters (m).
    
    \item \textbf{c\_walls}: Number of concrete and brick walls between the ED and GW, affecting signal attenuation.
    
    \item \textbf{w\_walls}: Number of wood partitions between the ED and GW, affecting signal attenuation.
    
    \item \textbf{exp\_pl}: Experimental path loss, a measure of signal attenuation over the distance between the transmitter (TX) and receiver (RX), in dB. It was calculated using the formula: \( 
        \mathrm{TP}_{\mathrm{TX}}\, \mathrm{(dBm)} 
        - \mathrm{CL}_{\mathrm{TX}}\, \mathrm{(dB)} 
        + G_{\mathrm{TX}}\, \mathrm{(dBi)} 
        + G_{\mathrm{RX}}\, \mathrm{(dBi)} 
        - \mathrm{CL}_{\mathrm{RX}}\, \mathrm{(dB)} 
        - \mathrm{RSSI}\, \mathrm{(dBm)}
            \).
    
    \item \textbf{n\_power}: NP, indicating the strength of the background interference affecting the received signal in dBm.
    
    \item \textbf{esp}: The ESP, quantifying the portion of the received power attributable to the actual signal, excluding background noise in dBm.
\end{enumerate}

\begin{figure*}[hbt!]
\centering
\centerline{\includegraphics[width=\textwidth]{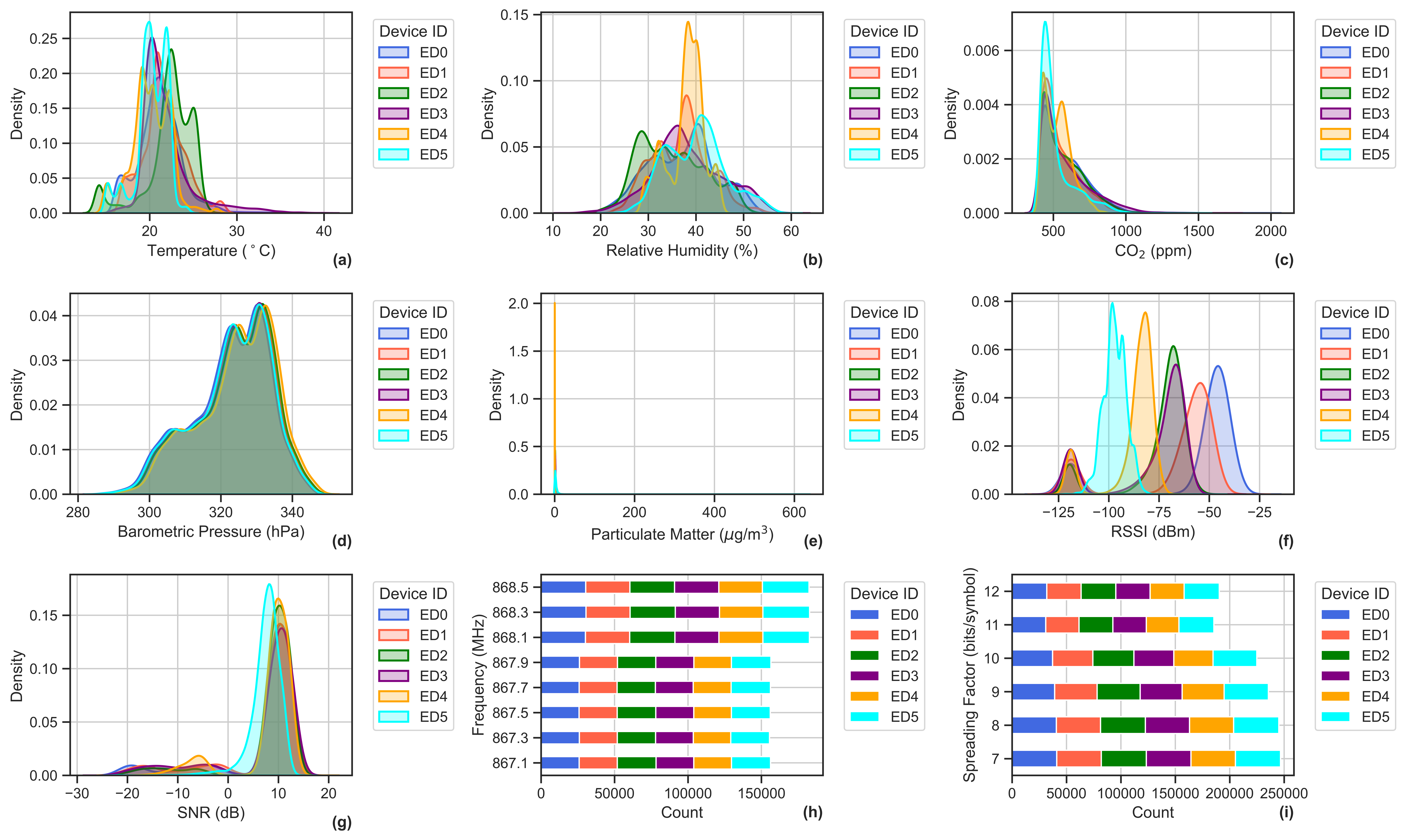}}
\caption{\label{fig:env_dist} {Empirical distributions of environmental and signal parameters affecting LoRaWAN performance across sensor nodes ED0 to ED5: (a) Temperature, (b) Relative Humidity, (c) CO\textsubscript{2}, (d) Barometric Pressure, (e) PM\textsubscript{2.5}, (f) RSSI, (g) SNR, (h) Frequency, and (i) Spreading Factor (SF).}}
\end{figure*}

As shown in Table~\ref{tab:stat_description}, the dataset reveals various environmental conditions and signal qualities within the indoor LoRaWAN setup. Notably, PM\textsubscript{2.5} levels exhibit significant variability, potentially reflecting fluctuations in human activity within the indoor environment. Signal metrics like RSSI and SNR show wide dispersions, indicating fluctuating network stability possibly influenced by factors such as environmental factors and physical barriers. Additionally, the distribution of SF and distances highlights the network's adaptability to varying transmission requirements. The data highlights the complexity of maintaining reliable LoRaWAN performance in dynamic indoor settings.

To help understand the variability and typical settings used in the deployed sensor network, the empirical distributions for environmental parameters that affect LoRaWAN performance are shown in Fig.~\ref{fig:env_dist}. Kernel density estimates (KDE) for key environmental factors: temperature, relative humidity, CO\textsubscript{2}, barometric pressure, and PM\textsubscript{2.5} are presented in  Fig.~\ref{fig:env_dist}(a) to Fig.~\ref{fig:env_dist}(e) in that order.

The temperature across the devices ranged from \( 13.74\,^\circ\text{C} \) to approximately \( 40.24\,^\circ\text{C} \). Devices ED1 and ED2 exhibited wider temperature variations, indicating a less controlled environment. This fluctuation could result from external factors like windows or ventilation affecting the indoor climate. In contrast, ED4, located in the server room, showed minimal variability in temperature, reflecting the controlled conditions of that environment.

Relative humidity varied between \( 13.99\% \) in the server room (ED4) and peaked around \( 60.19\% \) in the kitchen (ED5). The humidity distributions indicate that ED4 and ED3 maintain lower humidity levels, reflecting areas with less human activity or moisture-producing equipment. Higher humidity levels in ED5 are likely due to kitchen activities introducing moisture into the air.

CO\textsubscript{2} levels fluctuated from \( 386\,\mathrm{ppm} \) to \( 1993\,\mathrm{ppm} \). ED5, located \( 40\,\mathrm{m} \) away in the kitchen, recorded the highest CO\textsubscript{2} levels. This increase is likely a result of human activity and cooking processes, which elevate CO\textsubscript{2} concentrations.

Barometric pressure remained relatively stable between \( 286.91\,\mathrm{hPa} \) and \( 347.57\,\mathrm{hPa} \) across all devices. This consistency confirms the uniformity of the indoor environment on the same floor, indicating that localized environmental factors within the building less influence barometric pressure.

PM\textsubscript{2.5} concentrations were generally low across most devices, with a mean of \( 1.982\,\mathrm{\mu g/m^3} \), except in the kitchen (ED5), where elevated levels were observed. ED5 exhibited a right-skewed PM\textsubscript{2.5} distribution, likely resulting from microwave activity and associated particulate emissions. In contrast, ED4 showed minimal variability in PM\textsubscript{2.5}, indicative of the controlled environment within the server room.

Fig.~\ref{fig:env_dist}(f) and Fig.~\ref{fig:env_dist}(g) illustrate RSSI and SNR measurements, respectively. RSSI was highest (mean of \(-57.371\,\mathrm{dBm}\)) for ED0, positioned at the GW in the Hardware Lab, only \( 10\,\mathrm{m} \) away and separated by no walls, suggesting optimal signal reception due to its proximity and unobstructed line of sight. Conversely, ED5 in the kitchen, being the farthest from the GW at \( 40\,\mathrm{m} \) and separated by multiple walls, experienced the poorest mean RSSI (mean of \(-96.935\,\mathrm{dBm}\)) and a significantly the lowest mean SNR at SF12 (highest data rate)by approximately \(-5\,\mathrm{dB}\) compared to other devices. This degradation can be attributed to the increased distance and the signal attenuation caused by the combination of thin concrete, brick walls, and wood partitions, as well as potential interference from kitchen appliances. Additionally, ED2 and ED3 exhibited intermediate RSSI and SNR values, corresponding to their moderate distances and partial obstructions.

Analysis of frequency usage (Fig.~\ref{fig:env_dist}(h)) revealed greater utilization of the upper center frequency channels (\(867.9\,\mathrm{MHz}\) to \(868.5\,\mathrm{MHz}\)), possibly due to inherent channel-selection biases in device firmware, reduced interference levels, or gateway-side operational preferences. Regarding the spreading factor (Fig.~\ref{fig:env_dist}(i)), SF7 to SF10 were predominantly used, likely reflecting sufficient indoor link margins, shorter transmission distances typical of office environments, and stable propagation conditions. Despite disabled ADR and uniform cycling through all SFs, the reduced use of higher SFs (SF11–SF12) points to intrinsic optimization tendencies, potentially arising from the LoRaWAN stack or hardware-specific implementations. Nevertheless, further research is necessary to investigate these underlying mechanisms under diverse indoor scenarios systematically.

Overall, these findings emphasize the significant impact of environmental conditions and physical infrastructure on LoRaWAN network performance. Stable environments with minimal human activity, like the server room (ED4), contrast with dynamic areas, such as the kitchen (ED5), which experience more significant environmental fluctuations and higher human activity. This highlights the importance of assessing how different environmental factors influence the reliability of indoor wireless networks.

\begin{figure}[hbt!]
\centering
\centerline{\includegraphics[width=\columnwidth]{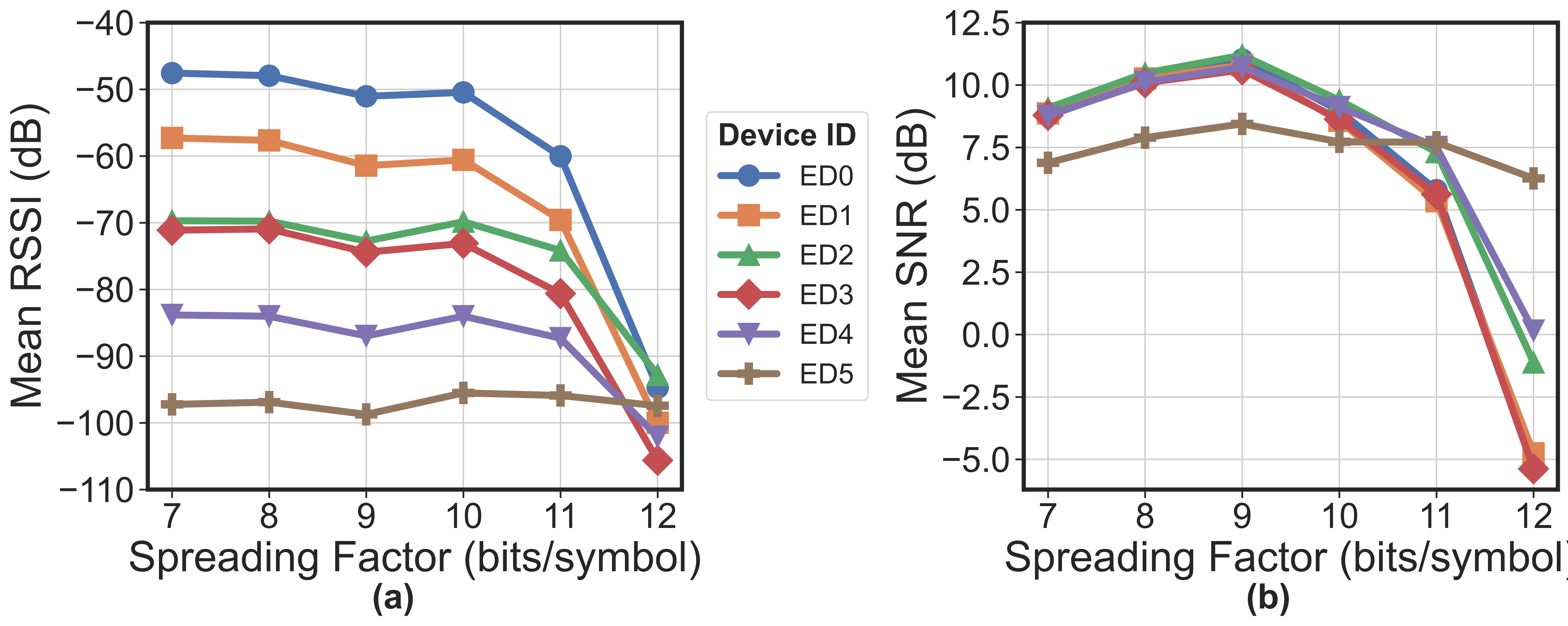}}
\caption{\label{fig: rssi_and_snr} {Mean RSSI and SNR across spreading factors (SFs) for indoor LoRaWAN deployments. (a) RSSI remains consistent up to SF10, with a notable decrease at SF11 and SF12. (b) SNR peaks between SF8 and SF10, sharply declining at SF12.}}
\end{figure}

Empirical results of mean RSSI and SNR in indoor LoRaWAN measurements at \(\mathrm{BW} = 125\,\mathrm{kHz}\) indicates diminished performance at SF11 and SF12 (Fig.~\ref{fig: rssi_and_snr}(a)), despite their theoretical link budget superiority. This arises from fundamental trade-offs inherent to LoRa modulation, amplified by indoor propagation dynamics: \textit{(i)} Extended symbol durations at SF11 and SF12 (\(T_{\text{symbol}} \approx 16.4\text{–}32.8\,\mathrm{ms}\)) exceed typical indoor channel coherence times (\(\sim 0.5\,\mathrm{ms}\)), causing phase incoherence during CSS demodulation. \textit{(ii)} In multipath-rich indoor environments, delayed signal replicas exceed the CSS's resilience window at SF11 and SF12 (\(T_{\text{symbol}} > 1\,\mathrm{s}\)), causing inter-symbol interference (ISI) that corrupts longer chirps. This is reflected in the nonlinear SNR degradation (Fig.~\ref{fig: rssi_and_snr}(b)), as coherent demodulation fails under time-varying channel conditions; \textit{(iii)} Prolonged ToA at high SFs increases exposure to transient interference (collision probabilities) in the shared 868 MHz band (e.g., microwave oven harmonics at \( 868\,\mathrm{MHz} \) in ED5's kitchen), a critical limitation in dense, NLoS indoor deployments characterized by heightened packet loss. Consequently, there is a need for adaptive network configurations and targeted optimizations, primarily in SF configuration, to enhance LoRaWAN performance across diverse indoor settings, ensuring reliable connectivity despite varying environmental and structural influences.

\begin{figure}[hbt!]
\centering
\centerline{\includegraphics[width=\columnwidth]{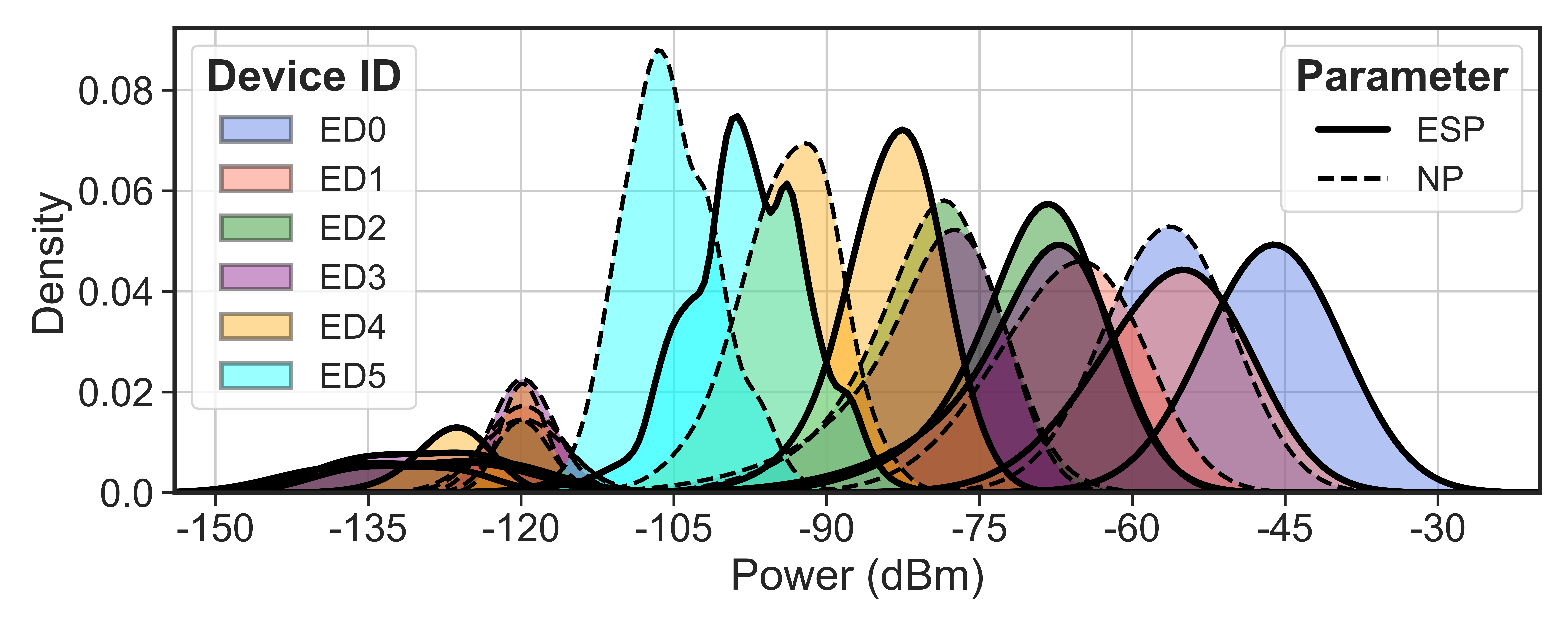}}
\caption{\label{fig:esp_n_power_comparison} {Density distributions of ESP and NP across all the devices (ED0 to ED5), illustrating LoRaWAN's resilience in indoor, high-interference environments.}}
\end{figure}

In indoor LoRaWAN environments, the ESP consistently remains above the NP, even in challenging propagation conditions. This characteristic ensures reliable communication across EDs, as illustrated by the density plots in Fig.~\ref{fig:esp_n_power_comparison}. Each device (ED0 through ED5) exhibits a distinct ESP distribution above its corresponding NP, denoted as \( \mathrm{ESP} > \mathrm{NP} \), where the difference \( \Delta \mathrm{SNR} = \mathrm{ESP} - \mathrm{NP} \) quantifies the SNR margin. This margin is crucial for sustaining connectivity in dense, obstacle-rich environments typical of indoor settings, where walls, floors, and electronic devices elevate NP. Such robustness highlights LoRaWAN's adaptability in NLoS and high-interference scenarios, allowing it to perform reliably in complex indoor spaces.

 Consequently, the PDRs for all EDs remain high and consistent, ranging from \( 88.82\% \) to \( 91.60\% \). Although ED0 is the nearest device, the minimal variation in PDR, just over \( 3\% \), demonstrates that the maintained SNR margins (\( \Delta \mathrm{SNR} \)) effectively ensure reliable communication throughout the indoor environment. This consistency highlights LoRaWAN's robustness in overcoming expected differences due to device placement, enabling dependable performance even in complex, obstacle-rich indoor settings. However, since the ADR was disabled to isolate fixed SF effects, this prevented devices from lowering their SF when \( \Delta \mathrm{SNR} \) exceeded thresholds, as witnessed by outliers in the distributions. For example, even when the SNR was high, devices on SF11 and SF12 had to keep using those slow settings, resulting in unnecessary exposure to interference and worse performance.

\section{Indoor Path Loss and Shadowing}
\label{sec: Indoor_Models}

This section builds on our EDA (Section~\ref{sec: data_description}) to quantify the effects of environmental parameters on indoor LoRaWAN propagation. It also serves as a practical example of how this dataset can be applied in signal modeling, link quality analysis, and performance evaluation.

\subsection{Log-Distance Path Loss and Shadowing Model with Multiple Walls Considerations}
\label{subsec: Indoor_Models}

The LDPLM is a fundamental framework used to predict signal attenuation as a function of distance \cite{goldsmithWirelessCommunications2005, azevedoCriticalReviewPropagation2024}. To enhance its applicability in complex indoor environments, this model is extended to account for multiple walls of varying materials and shadowing, thereby providing a more detailed assessment of signal degradation with the Log-Distance Path Loss Model with Shadowing and Multiple Walls(LDPLSM-MW). The path loss \( PL \) as a function of distance \( d \) and the number of walls \( W_k \) of different types \( k \) is expressed as given in  (\ref{eq:pl-mw}):
\begin{equation}
\label{eq:pl-mw}
\mathrm{PL}(d, \{W_k\}) = \mathrm{PL}(d_0) + 10n \times \log_{10}\left(\frac{d}{d_0}\right) + \sum_{k=0}^{K} W_k \times L_k + \epsilon,
\end{equation}
where \( \mathrm{PL}(d_0) \) is the path loss at the reference distance \( d_0 \), \( n \) is the path loss exponent, \( L_k \) is the attenuation per wall of type \( k \) in dB, \( W_k \) is the number of walls of type in the signal path, \( K \) is the total number of distinct wall types considered ( brick/concrete and wood in our case), \( d \) is the distance between the transmitter and receiver, and \( \epsilon \) is the shadowing term representing random signal fluctuations.

Shadowing accounts for random variations in path loss that deterministic models like distance-based attenuation and wall penetration losses do not capture. These fluctuations result from environmental factors such as reflections, scattering, and obstacles within indoor settings. In this work, we represent the shadowing term \( \epsilon \) in dB units, assuming it follows a Gaussian distribution with zero mean and variance \( \sigma_{\epsilon, \mathrm{dB}}^2 \), mathematically expressed as \( \epsilon \sim \mathcal{N}(0, \sigma_{\epsilon, \mathrm{dB}}^2) \), consistent with standard practice in log-distance path loss modeling. This Gaussian assumption effectively captures the stochastic nature of signal propagation, typically modeled by a log-normal distribution when converted to linear power scales.

When expressed on a linear scale, this shadowing component is commonly used to model the multiplicative effects of environmental variability on received signal strength. Its probability density function (PDF) is as given in  (\ref{eq: shadow_pdf}) \cite{goldsmithWirelessCommunications2005}.
\begin{equation}
\label{eq: shadow_pdf}
p(\epsilon) = \frac{\xi}{\sqrt{2\pi} \sigma_{\epsilon,\mathrm{dB}}\, \epsilon} 
\exp\left( -\frac{\left(10 \log_{10} \epsilon - \mu_{\epsilon,\mathrm{dB}} \right)^2}
{2 \sigma_{\epsilon,\mathrm{dB}}^2} \right), \quad \epsilon > 0,
\end{equation}
where \( \xi = \frac{10}{\ln 10} \) is a constant from the conversion between logarithmic and linear scales. The term \( \mu_{\epsilon,\mathrm{dB}} \) represents the mean of \( 10 \log_{10} \epsilon \), and \( \sigma_{\epsilon,\mathrm{dB}} \) is its standard deviation. The variable \( \epsilon \) denotes shadowing on the linear scale and it must be greater than zero.

\begin{figure}[hbt!]
\centering
\vspace{-10pt} 
\centerline{\includegraphics[width=.9\columnwidth]{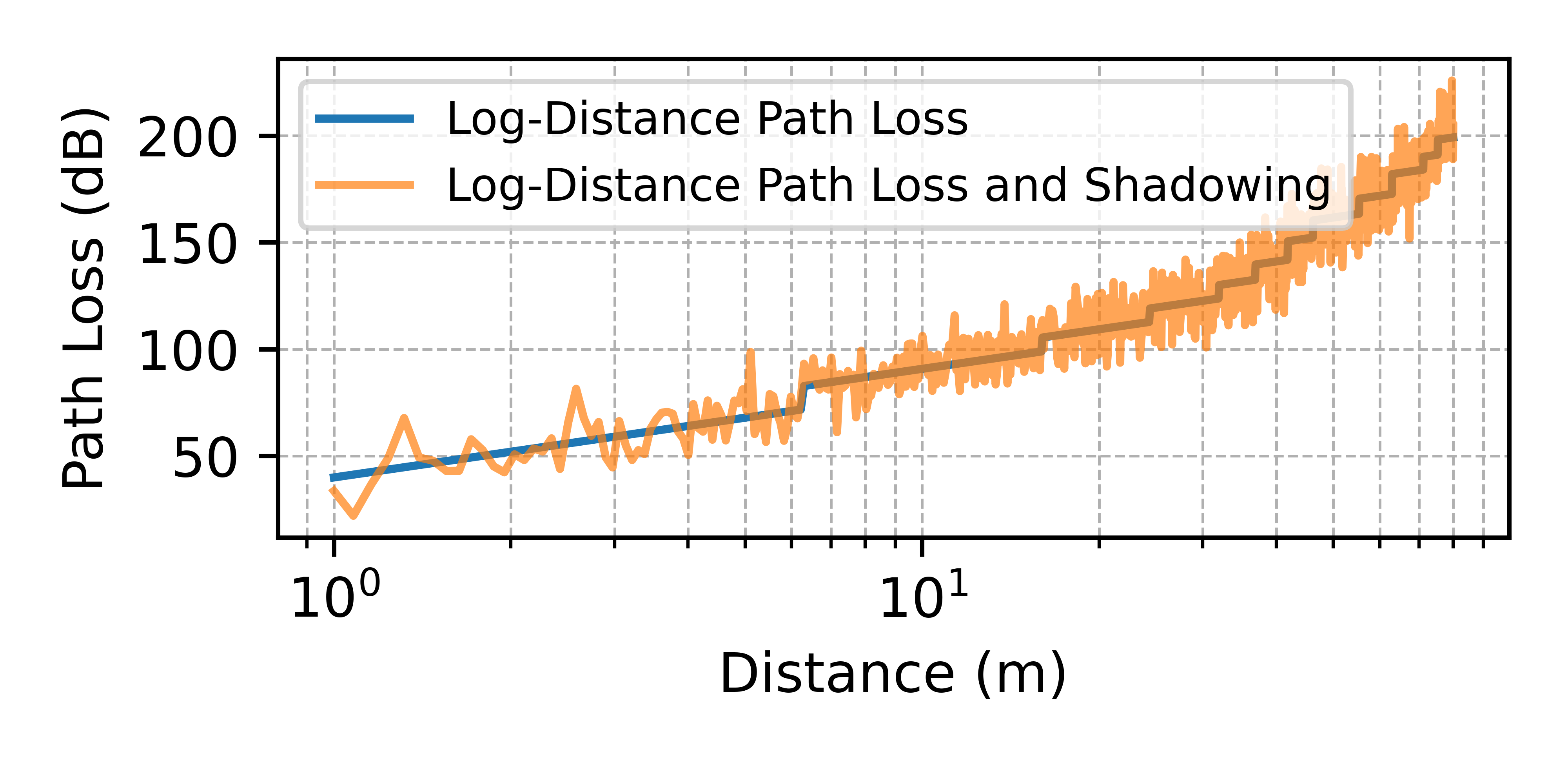}}
\vspace{-10pt} 
\caption{Simulated path loss versus distance in an indoor environment using the Log-Distance Path Loss Model with multi-wall effects and log-normal shadowing.}
\vspace{-10pt} 
\label{fig: path_loss_simulation}
\end{figure}

To illustrate the theoretical effects of path loss and shadowing in indoor environments, Fig.~\ref{fig: path_loss_simulation} presents a simulated path loss as a function of distance imposed with multiwall effects and log-normal shadowing. Simulation parameters are selected based on established literature over time \cite{goldsmithWirelessCommunications2005, obeidatIndoorPathLoss2018, azevedoCriticalReviewPropagation2024} to reflect realistic indoor conditions where the reference distance \( d_0 = 1 \, \mathrm{m} \) with \( \mathrm{PL}(d_0) = 40\,\mathrm{dB} \), the path loss exponent \( n = 3.5 \), wall attenuation \( L_{\text{wall}} \sim U(5, 12) \, \mathrm{dB} \) per wall, and the shadowing standard deviation \( \sigma_\epsilon = 9 \, \mathrm{dB} \). The elevated path loss exponent \( n = 4.0 \) typifies indoor settings where obstacles significantly degrade signals, unlike the free-space exponent \( n = 2 \). By modeling walls with random spacing \( s_{\text{wall}} \sim U(4, 10) \, \mathrm{m} \), we emulate typical building layouts and capture the cumulative impact of structural elements on signal propagation. This visualization effectively demonstrates how distance, multiple walls, and environmental variability contribute to indoor signal attenuation, highlighting the challenges in designing robust wireless communication systems.

\begin{figure*}[hbt!]
\begin{equation}
\label{eq:pl-mw-en}
\mathrm{PL}(d, \{W_k\}, \mathbf{E}) = \mathrm{PL}(d_0) + 10 \times n \times \log_{10}\!\Bigl(\tfrac{d}{d_0}\Bigr) + 20 \times \log_{10}(f) + \sum_{k=0}^{K} L_k \times W_k + \sum_{j=1}^{P} \theta_j \times E_j + k_{\text{SNR}} \times \mathrm{SNR} + \epsilon
\end{equation}
\hrulefill
\end{figure*}

\subsection{ Log-Distance Path Loss and Shadowing Model with Multiple Walls and  Environmental Parameters Considerations}

Building upon the foundational LDPLSM-MW, we propose an enhanced model that integrates additional environmental parameters (LDPLSM-MW-EP) to facilitate a more comprehensive and dynamic analysis of indoor signal propagation. This augmentation accounts for factors such as CO\textsubscript{2} concentration, relative humidity, PM\textsubscript{2.5} levels, barometric pressure, temperature, operating frequency, and the SNR, all of which have been found to influence LoRaWAN path loss significantly as discussed in the literature.

The LDPLSM-MW-EP path loss \( \mathrm{PL} \) is modeled as a function of distance \( d \), frequency \( f \), number of walls \( W_k \) of different types, and a vector of environmental parameters \( \mathbf{E} \) as given in~(\ref{eq:pl-mw-en}). The term \( 20 \times \log_{10}(f) \) accounts for the frequency-dependent component of the path loss, ensuring better accuracy for different operational frequencies. \( \mathbf{E} = [E_T, E_{\mathrm{RH}}, E_{\mathrm{BP}}, E_P, E_C] \) is the vector of environmental parameters—temperature (\( T \)), relative humidity (\( \mathrm{RH} \)), barometric pressure (\( \mathrm{BP} \)), PM\textsubscript{2.5} (\( P \)), and CO\textsubscript{2} (\( C \)). \( \theta_j \) are the coefficients corresponding to each environmental parameter \( E_j \), and \( k_{\mathrm{SNR}} \) is the scaling factor for the SNR. Lastly, \( \epsilon \) is the shadowing term whose probabilistic characterization enables the model to account for stochastic path-loss variations due to complex indoor environments.

\subsection{Log-Distance Path Loss and Shadowing Modeling}
This subsection gives the methodology for modeling indoor path loss using the LDPLSM-MW and the LDPLSM-MW-EP models and integrating the effects of multiple structural walls and various environmental parameters. Parameter estimation, performance evaluation, and cross-validation methodologies are also discussed to ensure robust and accurate modeling outcomes.

In our modeling approach, we represent the reference path loss at a distance \( d_0 \) as an empirically estimated intercept term \( \beta_0 \) instead of explicitly calculating \( \mathrm{PL}(d_0) \). This method directly captures fixed factors such as operating frequency (for the LDPLSM-MW model), antenna heights, and system-specific characteristics from the data, making the models more adaptable to real-world conditions. Also, by estimating \( \beta_0 \) from actual measurements, we avoid potential inaccuracies associated with theoretical assumptions and ensure that the baseline path loss accurately reflects our specific environment and system setup. This simplification enhances any model's clarity and predictive accuracy by focusing on the most influential factors affecting signal attenuation, such as environmental variables.

\subsubsection{Machine Learning Data Pipeline}

\begin{figure*}[hbt!]
\centering
\centerline{\includegraphics[width=0.8\textwidth]{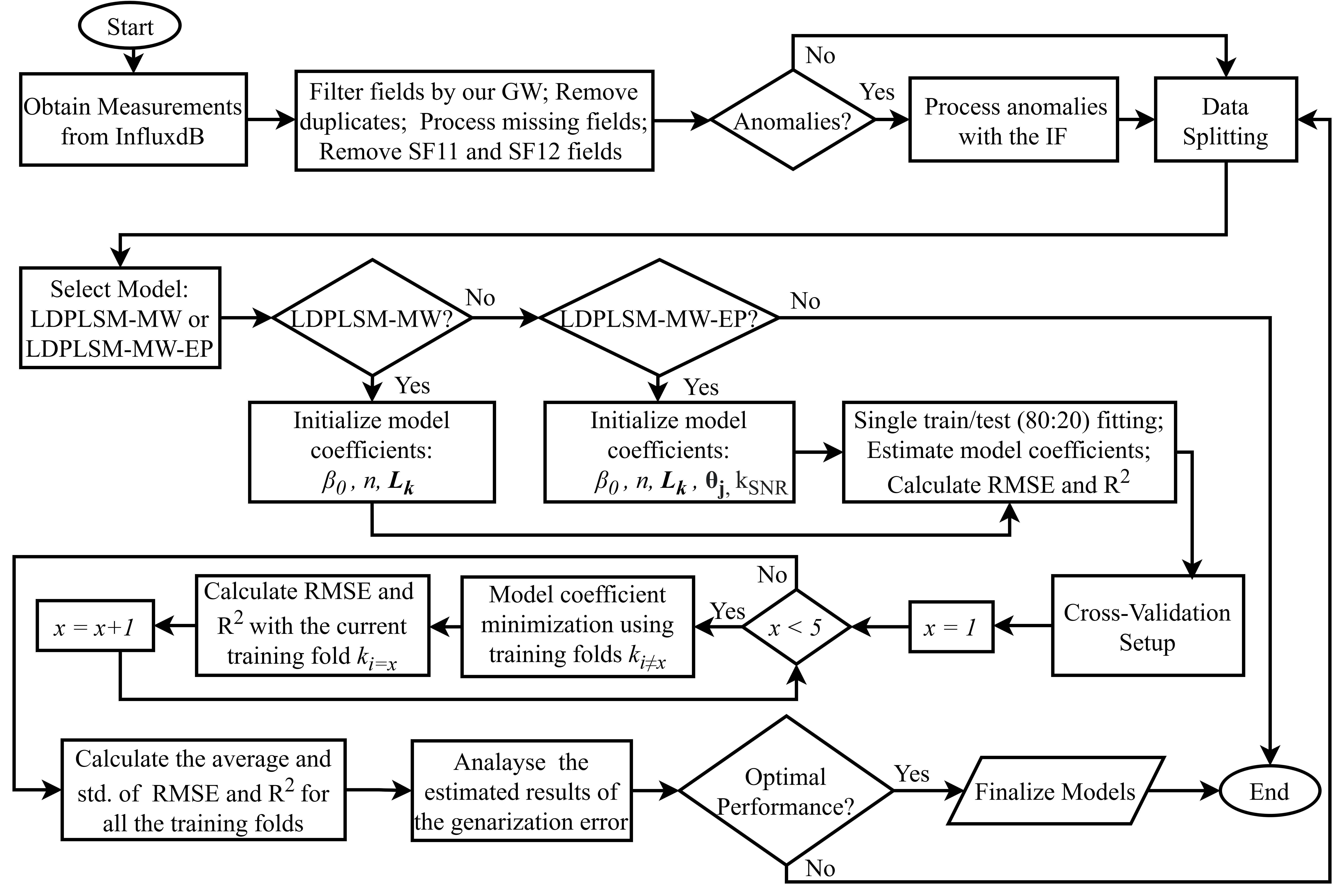}}
\caption{\label{fig: pipeline} {Data processing and evaluation pipeline, outlining steps from preprocessing and anomaly detection to cross-validation, optimization, and generalization error analysis.}}
\end{figure*}

Modeling the LDPLSM-MW and its enhanced variant of the LDPLSM-MW-EP for indoor environments necessitated a robust and precise data preparation pipeline, as illustrated in Fig.~\ref{fig: pipeline}. Utilizing the Python \texttt{pandas} library for data manipulation, we filtered the dataset to include only data from our deployed LoRaWAN GW, ensuring targeted and relevant analysis. For Class A LoRaWAN devices, retransmissions within short intervals are common due to acknowledgment and timing constraints. We grouped data per device to prevent skewness, sorted timestamps, and removed frames with unchanged counters within a 2-second window, ensuring consistent data and reliable modeling. Additionally, as justified under Fig.~\ref{fig: rssi_and_snr} and Fig.~\ref{fig:esp_n_power_comparison}, the SF11 and SF12 data were excluded from modeling to avoid bias from non-stationary channel effects, ensuring the generalizability of our LDPLSM in dynamic indoor IoT deployments. Subsequently, we conducted data cleaning by removing missing (NaN) and infinite values, accounting for approximately \(0.06\%\) of the dataset. These steps improved the dataset's consistency and validity for analysis.

\begin{figure}[hbt!]
\vspace{-5pt} 
\centering
\centerline{\includegraphics[width=0.9\columnwidth]{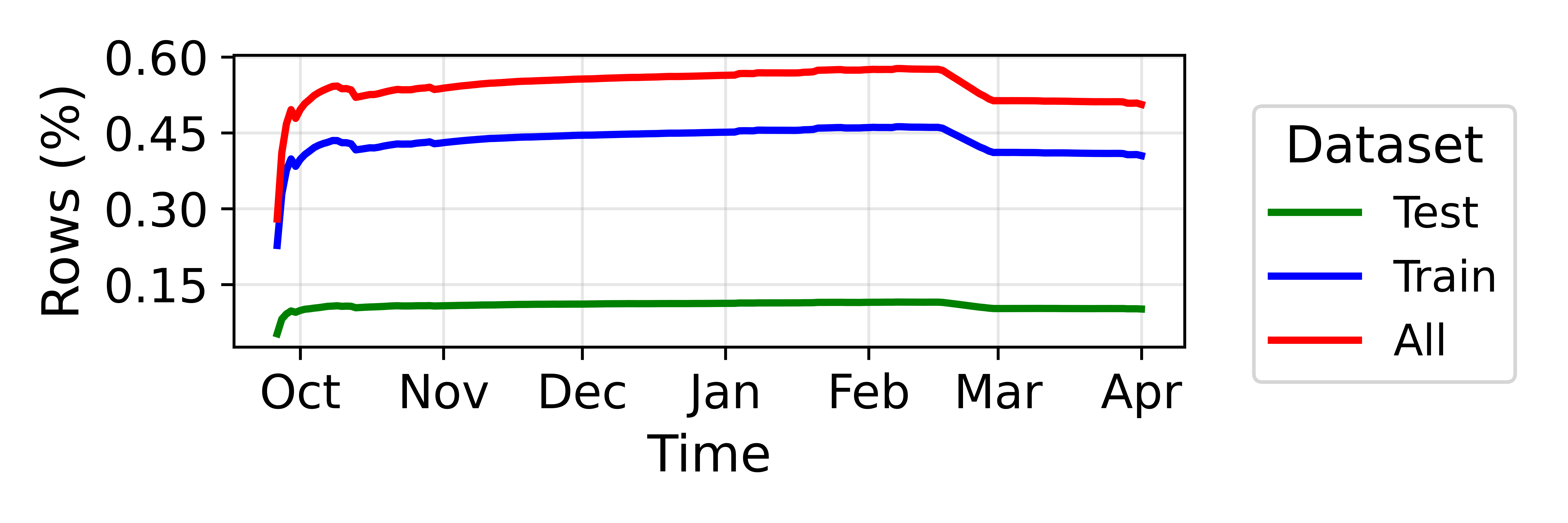}}
\caption{\label{fig: splitting} {Daily temporal distribution of observations for the entire dataset (All), training set (Train), and testing set (Test). Percentages are smoothed with a rolling mean to show trends, ensuring consistent subset representativeness.}}
\vspace{-5pt} 
\end{figure}

After data cleaning, we employed the Isolation Forest (IF) algorithm for anomaly detection from the \texttt{scikit-learn} library. This method was applied individually to each ED to capture device-specific operational patterns. We focused on seven key sensor features: CO\textsubscript{2} levels, relative humidity, PM\textsubscript{2.5}, barometric pressure, temperature, RSSI, and SNR. Based on a related study \cite{gonzalez-palacioLoRaWANPathLoss2023}, we set our Isolation Forest’s \texttt{contamination} parameter to \(0.01\), relaxing the strict \(0.1\%\) threshold used in the Mahalanobis method (\(p = 0.001\)), to better align with the complex multipath and noise effects in high-dynamic indoor environments. We utilized the \texttt{StandardScaler} function to ensure consistent feature scaling. To guarantee reproducibility, the IF was configured with 100 estimators and a \texttt{random\_state} of 42. The model identified and removed anomalies by processing data across all devices, enhancing data integrity and ensuring reliable training for predictive models.
indicating that approximately  of the data might be anomalous,

The refined dataset was partitioned into training and testing subsets using an 80-20 split, a common approach in modeling \cite{chopraIntroductionMachineLearning2023}. This was achieved using the \texttt{train\_test\_split} function, which also employed uniformly distributed random split as utilized by other PLM studies \cite{gonzalez-palacioMachineLearningBasedCombinedPath2023a}. The split was validated by computing the percentage of rows per subset over time, resampled daily to maintain consistency across temporal windows. This strategy ensured that the distribution of key predictor variables remained consistent across both subsets, maintaining the dataset's representativeness. Additionally, a fixed random seed was set to guarantee reproducibility, which is essential for validating and comparing model performance across different iterations. Fig.~\ref{fig: splitting} displays the daily distribution of observations for the full dataset and its train/test subsets. Each day accounts for roughly \(0.6\%\) of the total data, and the rolling mean highlights only minor fluctuations over time. This consistent distribution is critical for accurately capturing the complex indoor environmental dynamics and structural configurations, thereby ensuring robust and representative predictive modeling.

We utilized the parameters described in (\ref{eq:pl-mw-en}) to accurately capture the complex interplay between structural and environmental factors affecting indoor signal propagation. This involved specifying the number and types of walls in the signal path and incorporating data from office-installed sensors to account for temporal variables like time of day and occupancy patterns. Focusing on these critical predictors enhanced our models' efficiency and clarity without using traditional dimensionality reduction techniques. During model training, we initialized the base model weights (\( \beta_0 \), \( n \), and \( L_k \)) based on domain expertise as simulated in Section.~\ref{subsec: Indoor_Models}. The environmental coefficients (\( \theta_j \)) were initialized to \(0.0\), reflecting a conservative assumption of negligible direct impact unless supported by the data, thereby avoiding premature bias in the model. The SNR scaling factor (\( k_{\text{SNR}} \)) was initialized to \(-1.0\), consistent with the inverse relationship between SNR and path loss observed in preliminary analyses. Evaluation on a reserved test dataset showed strong performance, with the RMSE and \(R\textsuperscript{2}\) calculated using the \texttt{scikit-learn} library metrics confirming high accuracy and the models' ability to explain data variability effectively. This approach successfully modeled the intricate dynamics of indoor propagation environments, demonstrating the strengths of both LDPLSM-MW and LDPLSM-MW-EP in capturing the key factors that influence LoRaWAN signal propagation.

\begin{table*}[hbt!]
\centering
\caption{Predictor Variable Coefficients and Comparison of Performance Metrics for the LDPLSM-MW and LDPLSM-MW-EP Models}
\label{tab:modelcoeffs}
\begin{tabular}{ccccc}
\hline
\textbf{Parameter}  & \textbf{Unit} & \textbf{Model Variable} & \textbf{LDPLSM-MW} & \textbf{LDPLSM-MW-EP} \\ 
\hline
Intercept  & dB & $\beta_0$ & 31.30 & 5.46 \\ 
Path loss exponent  & - & $n$ & 3.62 & 3.20 \\ 
Brick Wall Loss  & dB & $L_{\text{brick}}$ & 9.74 & 8.52 \\ 
Wood Wall Loss & dB & $L_{\text{wood}}$ & 2.64 & 2.98 \\ 
CO$_{2}$  & dB/ppm & $\theta_{\text{C}}$ & - & -0.002497 \\ 
Relative humidity & dB/\% & $\theta_{\text{RH}}$ & - & -0.074299 \\ 
PM$_{2.5}$  & dB/($\mu$g/m$^3$) & $\theta_{\text{P}}$ & - & -0.153205 \\
Barometric pressure & dB/hPa & $\theta_{\text{BP}}$ & - & -0.011567 \\ 
Temperature  & dB/\textdegree C & $\theta_{\text{T}}$ & - & -0.005767 \\ 
SNR scaling factor & - & $k_{\text{SNR}}$ & - & -1.982231 \\ 
\hline
\end{tabular}
\end{table*}

To enhance model robustness and generalizability, we implemented a 5-fold cross-validation strategy using the \texttt{KFold} class. The training data was divided into five folds, \(k_1, k_2, k_3, k_4,\) and \(k_5\). For each fold \(k_i\) where \(i = 1, 2, 3, 4, 5\), the model was trained on the remaining four folds and validated on \(k_i\). The RMSE was calculated for each iteration using the \texttt{mean\_squared\_error} function, and the average RMSE along with its standard deviation were computed to assess model performance as per \cite{chopraIntroductionMachineLearning2023}. Models exhibiting low average RMSE and minimal standard deviation were deemed to possess strong generalization capabilities, effectively mitigating overfitting and underfitting risks.

\subsubsection{Parameter Estimation}

Accurate parameter estimation is crucial for aligning the theoretical models with empirical data, ensuring that the models accurately reflect observed signal attenuation patterns. The LDPLSM-MW and LDPLSM-MW-EP models employ non-linear curve fitting techniques to estimate the unknown parameters, achieving high fidelity between predicted and measured path loss values.

The primary objective in parameter estimation is to minimize the discrepancy between the predicted path loss (\( \mathrm{PL}_{\mathrm{pred}} \)) and the experimentally measured path loss (\( \mathrm{PL}_{\mathrm{exp}} \)). This was achieved by minimizing the residual sum of squares (RSS) given in (\ref{eq: rss}):
\begin{equation}
\label{eq: rss}
\mathrm{RSS}(\boldsymbol{\alpha}) = \sum_{i=1}^{N} \left(\mathrm{PL}_{\mathrm{exp},i} - \mathrm{PL}_{\mathrm{pred},i}(\boldsymbol{\alpha})\right)^2,
\end{equation}
where \(\boldsymbol{\alpha}\) represents the vector of model parameters: (\(\beta_0, n, L_k, \theta_{j}\)) for the LDPLSM-MW and (\(\beta_0, n, L_k, \theta_{j}, k_{\mathrm{SNR}}\)) for the LDPLSM-MW-EP. \(N\) is the total number of observations,\( \mathrm{PL}_{\mathrm{exp},i} \) is the experimentally measured path loss for the \( i \)-th observation, and \( \mathrm{PL}_{\mathrm{pred},i} \) is the predicted path loss for the \(i\)-th observation based on the model: (\ref{eq:pl-mw}) or (\ref{eq:pl-mw-en}).

To solve this non-linear optimization problem, the Levenberg-Marquardt algorithm was employed, effectively blending gradient descent and the Gauss-Newton method to ensure efficient and stable convergence. The iterative update rule is given by (\ref{eq: iterative}):
\begin{equation}
\boldsymbol{\alpha}_{m+1} = \boldsymbol{\alpha}_m - \left(J^T J + \lambda I\right)^{-1} J^T \mathbf{r},
\label{eq: iterative}
\end{equation}
where \( J \) is the Jacobian matrix of partial derivatives \( \frac{\partial \mathrm{PL}_{\mathrm{pred}}}{\partial \alpha_j} \), \( \mathbf{r} \) is the residual vector \( (\mathrm{PL}_{\mathrm{exp}} - \mathrm{PL}_{\mathrm{pred}}) \), \(\lambda\) is the damping factor, and \(I\) is the identity matrix. This approach ensures robust parameter updates, balancing the convergence speed and stability by dynamically adjusting \(\lambda\) to mediate between the gradient descent and Gauss-Newton methods.

The parameter estimation was performed using \texttt{SciPy}'s \texttt{curve\_fit} function, which leverages the Levenberg-Marquardt algorithm for non-linear curve fitting. This sophisticated iterative method combines the strengths of gradient descent and the Gauss-Newton approach to minimize the RSS efficiently. To initiate the optimization, informed initial guesses for the model parameters \(\boldsymbol{\alpha}\) were provided, facilitating quicker convergence. The algorithm then iteratively refined these parameters, systematically reducing the RSS through up to \(100{,}000\) iterations or until the RSS stabilized below a predefined threshold, ensuring robust and reliable convergence. 

After optimization, the fitted parameters were rigorously validated to ensure they were physically meaningful and statistically significant. This validation included assessing the parameters' plausibility within the context of the modeled environment and evaluating their statistical robustness. Thus, the parameter estimates reliably predicted unseen data and accurately reflected the underlying signal attenuation phenomena.

\begin{figure*}[hbt!]
\centering
\centerline{\includegraphics[width=\textwidth]{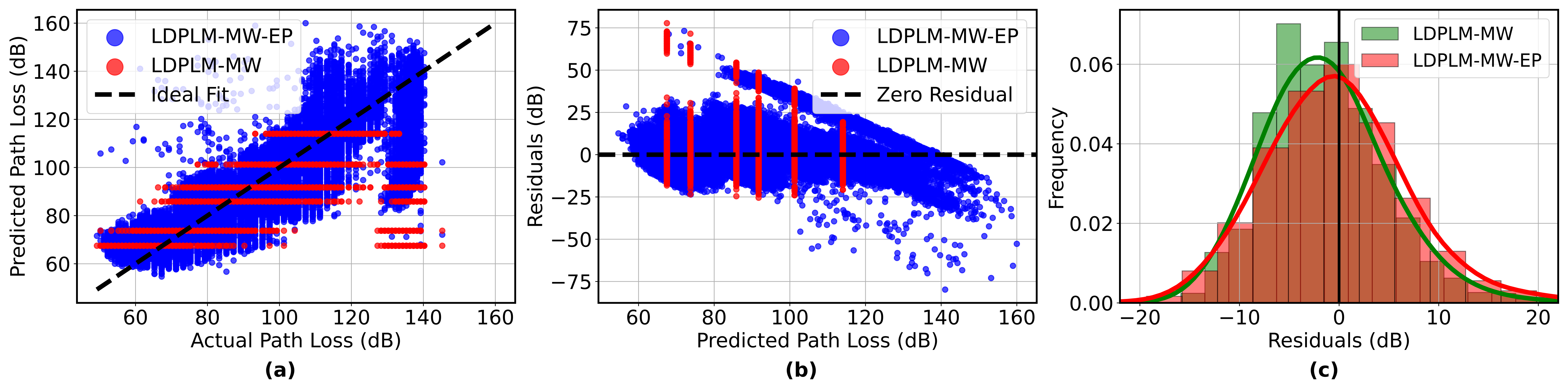}}
\caption{\label{fig: models_evaluation} Evaluation of the LDPLSM-MW and LDPLSM-MW-EP models in predicting path loss: (a) Predicted vs. actual path loss with an ideal fit line for comparison; (b) residuals showing deviation from zero error; (c) histogram of residuals illustrating error distribution for each model.}
\end{figure*}

\subsubsection{Analysis and Discussion}

The model coefficients shown in Table~\ref{tab:modelcoeffs} provide insights into indoor signal propagation. First, the path loss exponents are approximately \(3.62\) and \(3.20\), falling within the typical indoor range of \(2\) to \(7\) \cite{azevedoCriticalReviewPropagation2024}. This indicates that signals weaken more quickly inside buildings than in free space due to factors like reflections, scattering, and obstacles. Second, wall losses are approximately \(9.0\,\mathrm{dB}\) for brick and between \(2.64\,\mathrm{dB}\) to \(2.98\,\mathrm{dB}\) for wood, consistent with literature values and emphasizing the higher attenuation caused by denser materials \cite{goldsmithWirelessCommunications2005}. Third, while outdoor studies typically find that the path loss has a direct proportional relation with environmental parameters \cite{gonzalez-palacioLoRaWANPathLoss2023}, our indoor model shows negative coefficients. This likely reflects the complex interplay of factors in indoor environments; for example, higher humidity might coincide with open doors, furniture being moved to create a LoS, or reduced occupancy that leads to fewer human-induced obstructions. The SNR coefficient was also negative, meaning that lower SNR correlates with higher path loss, as expected \cite{gonzalez-palacioLoRaWANPathLoss2023}. These environmental parameters may act as proxies for other indoor factors affecting signal propagation. Overall, indoor signal behavior is shaped by distance, wall types, and complex environmental interactions.

Fig.~\ref{fig: models_evaluation} compares the performance of the LDPLSM-MW model with its enhanced version, LDPLSM-MW-EP, highlighting the significant impact of incorporating environmental parameters into path loss predictions. In Fig.~\ref{fig: models_evaluation}(a), the scatter plot of predicted versus actual path loss shows that LDPLSM-MW-EP predictions align closely with the line of perfect agreement, indicating superior predictive accuracy over LDPLSM-MW. Fig.~\ref{fig: models_evaluation}(b) illustrates the residuals for both models; LDPLSM-MW-EP exhibits more minor deviations around zero, reflecting reduced bias and an improved overall model fit. Fig.~\ref{fig: models_evaluation}(c) reveals that, although the LDPLSM-MW model has a sharper peak near zero, its residual distribution is less symmetric compared to that of the LDPLSM-MW-EP model. The LDPLSM-MW-EP model shows a more balanced distribution around zero, indicating that prediction errors are evenly spread on both sides. This symmetry suggests a reduction in systematic bias, highlighting the model's more reliable performance despite a marginally broader central spread.

As presented in Table~\ref{table: average_metrics}, quantitative analysis of the model residuals reveals that the mean residual error shifted from \(-0.0059\,\mathrm{dB}\) for the LDPLSM-MW model to \(0.0031\,\mathrm{dB}\) for the LDPLSM-MW-EP model. Although this change indicates a slight directional shift from negative to positive bias, the magnitude remains negligible, thus maintaining acceptable accuracy levels for practical PLM applications. More notably, the skewness of the residual distribution markedly decreased from \(3.7250\) to \(1.0746\), signifying a substantial improvement towards symmetry. This implies a lower frequency of extreme prediction errors, enhancing the consistency and reliability of predictions from the LDPLSM-MW-EP model.

In our analysis, we quantitatively assessed the performance of the LDPLSM-MW and LDPLSM-MW-EP models using a standard train-test split, as summarized in Table~\ref{table:performance_metrics}. The enhanced LDPLSM-MW-EP model exhibited substantial improvements over the original LDPLSM-MW model across all evaluated metrics. Specifically, the RMSE on the training set decreased from \(10.5773\,\mathrm{dB}\) to \(8.0343\,\mathrm{dB}\), while on the test set it decreased from \(10.5771\,\mathrm{dB}\) to \(8.0343\,\mathrm{dB}\). This notable reduction demonstrates the effectiveness of incorporating environmental parameters in improving path loss prediction accuracy.

\begin{table}[hbt!]
  \centering
  \caption{Performance metrics for the LDPLSM-MW and the LDPLSM-MW-EP models, including RMSE, R\textsuperscript{2}, and shadowing error across training and testing subsets.}
  \label{table:performance_metrics}
    \begin{tabular}{
      >{\centering\arraybackslash}m{1.7cm} 
      >{\centering\arraybackslash}m{0.7cm} 
      >{\centering\arraybackslash}m{1.5cm} 
      >{\centering\arraybackslash}m{0.4cm} 
      >{\centering\arraybackslash}m{2.2cm}
    } 
    \hline
    \textbf{Model} & \textbf{Subset} & \textbf{RMSE (dB)} & R\textsuperscript{2} & \textbf{Shadowing $\boldsymbol{\sigma}$ (dB)} \\ \hline
    \multirow{2}{*}{LDPLSM-MW} 
                     & Train & 10.5786 & 0.6913 & \multirow{2}{*}{10.5786} \\ 
                     & Test  & 10.5721 & 0.6917 & \\ \hline
    \multirow{2}{*}{LDPLSM-MW-EP} 
                     & Train & \textbf{8.0357}  & \textbf{0.8219} & \multirow{2}{*}{\textbf{8.0357}} \\  
                     & Test  & \textbf{8.0290}  & \textbf{0.8222} & \\ \hline
    \end{tabular}
\end{table}

The values of R\textsuperscript{2} showed parallel improvements, increasing from \( 0.6914 \) to \( 0.8219 \) on the training set and remaining consistent at \( 0.6914 \) to \( 0.8219 \) on the test set. These higher R\textsuperscript{2} values indicate that the LDPLSM-MW-EP model accounts for a greater proportion of variance in the measured path loss, making its predictions more reliable. Furthermore, the consistency between training and testing metrics highlights the enhanced model’s ability to generalize effectively to unseen data, mitigating the risk of overfitting.

The results of the 5-fold cross-validation are summarized in Table~\ref{table: average_metrics}. During this evaluation, the average RMSE on the training sets improved significantly, decreasing from \( 10.5773 \pm 0.0166\,\mathrm{dB} \) for the LDPLSM-MW model to \( 8.0343 \pm 0.0096\,\mathrm{dB} \) for the LDPLSM-MW-EP model, where the \( \pm \) values indicate the standard deviations across folds. Similarly, the RMSE on the test sets showed a marked reduction, dropping from \( 10.5771 \pm 0.0666\,\mathrm{dB} \) to \( 8.0343 \pm 0.0382\,\mathrm{dB} \), highlighting consistent performance enhancements across all folds.

\begin{table}[hbt!]
\centering
\caption{Quantitative residual error distribution and average metrics (as an estimate of generalization error) across 5-fold cross-validation (\( \pm \) indicates the standard deviation across the folds)}
\label{table: average_metrics}
\begin{tabular}{cccc}
\hline
\textbf{Metric} & \textbf{Subset} & \textbf{LDPLSM-MW} & \textbf{LDPLSM-MW-EP} \\ \hline
Mean  (dB)  & Test  & \(-0.0059\)      & \(0.0031\)    \\ 
Skewness   & Test  & \(3.7250\)       & \(1.0746\)     \\ 
RMSE (dB)   & Train  & \(10.5773 \pm 0.0166\)   & \(8.0343 \pm 0.0096\)  \\ 
RMSE (dB)    & Test  & \(10.5771 \pm 0.0666\)   & \(8.0343 \pm 0.0382\)  \\ 
R\textsuperscript{2} & Train & \(0.6914 \pm 0.0008\)  & \(0.8219 \pm 0.0003\)   \\ 
R\textsuperscript{2} & Test & \(0.6914 \pm 0.0031\)  & \(0.8219 \pm 0.0012\)  \\ 
\hline
\end{tabular}
\end{table}

R\textsuperscript{2} values during cross-validation also improved significantly. For the LDPLSM-MW-EP model, the average of R\textsuperscript{2} on the training sets increased from \( 0.6914 \pm 0.0008 \) to \( 0.8219 \pm 0.0003 \), while on the test sets it increased from \( 0.6914 \pm 0.0031 \) to \( 0.8219 \pm 0.0012 \). These enhanced metrics demonstrate the model's improved capacity to explain variance in path loss measurements, ensuring reliability and robustness across varying data subsets.

These quantitative enhancements demonstrate that incorporating environmental parameters into the PLM reduces prediction errors and improves the model's ability to generalize across different indoor environments. Therefore, the LDPLSM-MW-EP model offers a more robust and accurate tool for predicting indoor path loss, which is critical for designing and optimizing wireless communication systems in complex environments.

\section{Future Research Directions}

To further advance our understanding, we plan to scale up our data collection campaign to multiple floors and diverse indoor settings to enrich the dataset. We will also explicitly analyze and visualize the spatial-temporal relationship between path loss and distance between the EDs and the GW.  To analyze the influence of environmental parameters such as floor level, building materials, and occupancy, we will conduct statistical significance tests, including Analysis of Variance (ANOVA) and t-tests for parametric analysis, as well as Kruskal Wallis and Mann Whitney tests for non-parametric scenarios. For modeling path loss, we will also employ non-parametric regression methods like Gaussian Process Regression, Random Forests, and Gradient Boosting alongside neural networks to capture complex relationships and improve predictive accuracy. Further enhancements will focus on integrating domain-specific features into various architectures, employing advanced hyperparameter optimization techniques, and utilizing robust cross-validation methods to ensure reliable generalization across varied environments.

\section{Conclusion}
\label{sec: conclusion}

This study leveraged a large-scale dataset of LoRaWAN transmissions comprising \( 1,\!328,\!334 \) fields collected in a real-world office environment at the University of Siegen’s Campus Hölderlinstraße, using readily available off-the-shelf components. The dataset does not just capture traditional propagation metrics like RSSI and SNR but also includes a spectrum of environmental parameters, such as temperature, humidity, CO\textsubscript{2}, barometric pressure, and PM\textsubscript{2.5}, providing a rich foundation for advancing indoor IoT communication models.

First, by examining a baseline LDPLSM-MW, we established how well classical approaches can predict attenuation across various structural layouts. We then introduced our enhanced LDPLSM-MW-EP model, incorporating these dynamic environmental factors and SNR. This enrichment offered a marked improvement in predictive accuracy, lowering the RMSE from about \(10.58\,\mathrm{dB}\) to \(8.04\,\mathrm{dB}\) and boosting R\textsuperscript{2} from approximately \(0.6914\) to \(0.8219\). In other words, modeling the subtleties of indoor conditions, factors as diverse as wall compositions and ambient environmental conditions, translate into substantially better path loss predictions.

These findings show that even within the complexity of indoor settings, an informed approach that integrates environmental insights can transform planning and design strategies for LoRaWAN networks. The results validate the feasibility of such approaches with standard commercial hardware and highlight a new frontier where data-rich environmental models help stakeholders make more confident decisions about GW placements, node density, and network parameters. As indoor IoT deployments grow increasingly intricate, our expanded dataset and refined modeling framework can serve as valuable touchstones, guiding practitioners and researchers toward more robust, context-aware communication solutions.

\section*{Data Availability Statement}
The dataset collected during the indoor LoRaWAN measurement campaign is available on GitHub at \url{https://github.com/nahshonmokua/LoRaWAN-Indoor-PathLoss-Dataset-IEEEACCESS} (accessed on 8~May~2025). The repository also includes a full script suite for device firmware flashing, data retrieval, preprocessing, and analysis used in this work.

\section*{Acknowledgment}
The authors thank the German Academic Exchange Service (DAAD) and the University of Siegen for their support and the Ubiquitous Computing Research Group for their valuable assistance during the data collection campaign.

\balance

\bibliography{refs}

\end{document}